\documentclass[twocolumn,traditabstract]{aa}

\usepackage[nonamebreak]{natbib}
\usepackage[stable]{footmisc}
\usepackage[fleqn]{amsmath} 
\usepackage{txfonts}
\usepackage{placeins}
\usepackage{natbib}
\bibpunct{(}{)}{;}{a}{}{,}
\usepackage{graphicx}
\usepackage{epstopdf}
\usepackage{ifthen}
\usepackage{mathtools}
\usepackage[breaklinks, colorlinks, citecolor=blue]{hyperref}
\usepackage{fixltx2e}
\usepackage{url}
\usepackage[table,usenames,dvipsnames]{xcolor}
\usepackage{ulem}

\def\setsymbol#1#2{\expandafter\def\csname #1\endcsname{#2}}
\def\getsymbol#1{\csname #1\endcsname}

\def\Planck{\textit{Planck}}

\newbox\tablebox    \newdimen\tablewidth
\def\leaderfil{\leaders\hbox to 5pt{\hss.\hss}\hfil}
\def\endPlancktablewide{\tablewidth=\textwidth 
    $$\hss\copy\tablebox\hss$$
    \vskip-\lastskip\vskip -2pt}
\def\tablenote#1 #2\par{\begingroup \parindent=0.8em
    \abovedisplayshortskip=0pt\belowdisplayshortskip=0pt
    \noindent
    $$\hss\vbox{\hsize\tablewidth \hangindent=\parindent \hangafter=1 \noindent
    \hbox to \parindent{$^#1$\hss}\strut#2\strut\par}\hss$$
    \endgroup}
\def\doubleline{\vskip 3pt\hrule \vskip 1.5pt \hrule \vskip 5pt}

\def\L2{\ifmmode L_2\else $L_2$\fi}

\def\DeltaT{\ifmmode \Delta T\else $\Delta T$\fi}
\def\deltat{\ifmmode \Delta t\else $\Delta t$\fi}
\def\fknee{\ifmmode f_{\rm knee}\else $f_{\rm knee}$\fi}
\def\Fmax{\ifmmode F_{\rm max}\else $F_{\rm max}$\fi}
\def\solar{\ifmmode{\rm M}_{\mathord\odot}\else${\rm M}_{\mathord\odot}$\fi}
\def\Msolar{\ifmmode{\rm M}_{\mathord\odot}\else${\rm M}_{\mathord\odot}$\fi}
\def\Lsolar{\ifmmode{\rm L}_{\mathord\odot}\else${\rm L}_{\mathord\odot}$\fi}
\def\inv{\ifmmode^{-1}\else$^{-1}$\fi}
\def\mo{\ifmmode^{-1}\else$^{-1}$\fi}
\def\sup#1{\ifmmode ^{\rm #1}\else $^{\rm #1}$\fi}
\def\expo#1{\ifmmode \times 10^{#1}\else $\times 10^{#1}$\fi}
\def\,{\thinspace}
\def\lsim{\mathrel{\raise .4ex\hbox{\rlap{$<$}\lower 1.2ex\hbox{$\sim$}}}}
\def\gsim{\mathrel{\raise .4ex\hbox{\rlap{$>$}\lower 1.2ex\hbox{$\sim$}}}}

\def\simprop{\mathrel{\raise .4ex\hbox{\rlap{$\propto$}\lower 1.2ex\hbox{$\sim$}}}}
\def\deg{\ifmmode^\circ\else$^\circ$\fi}
\def\pdeg{\ifmmode $\setbox0=\hbox{$^{\circ}$}\rlap{\hskip.11\wd0 .}$^{\circ}
          \else \setbox0=\hbox{$^{\circ}$}\rlap{\hskip.11\wd0 .}$^{\circ}$\fi}
\def\arcs{\ifmmode {^{\scriptstyle\prime\prime}}
          \else $^{\scriptstyle\prime\prime}$\fi}
\def\arcm{\ifmmode {^{\scriptstyle\prime}}
          \else $^{\scriptstyle\prime}$\fi}
\newdimen\sa  \newdimen\sb
\def\parcs{\sa=.07em \sb=.03em
     \ifmmode \hbox{\rlap{.}}^{\scriptstyle\prime\kern -\sb\prime}\hbox{\kern -\sa}
     \else \rlap{.}$^{\scriptstyle\prime\kern -\sb\prime}$\kern -\sa\fi}
\def\parcm{\sa=.08em \sb=.03em
     \ifmmode \hbox{\rlap{.}\kern\sa}^{\scriptstyle\prime}\hbox{\kern-\sb}
     \else \rlap{.}\kern\sa$^{\scriptstyle\prime}$\kern-\sb\fi}
\def\ra[#1 #2 #3.#4]{#1\sup{h}#2\sup{m}#3\sup{s}\llap.#4}
\def\dec[#1 #2 #3.#4]{#1\deg#2\arcm#3\arcs\llap.#4}
\def\deco[#1 #2 #3]{#1\deg#2\arcm#3\arcs}
\def\rra[#1 #2]{#1\sup{h}#2\sup{m}}

\def\dots{\relax\ifmmode \ldots\else $\ldots$\fi}
\def\WHzsr{\ifmmode $W\,Hz\mo\,sr\mo$\else W\,Hz\mo\,sr\mo\fi}
\def\mHz{\ifmmode $\,mHz$\else \,mHz\fi}
\def\GHz{\ifmmode $\,GHz$\else \,GHz\fi}
\def\mKs{\ifmmode $\,mK\,s$^{1/2}\else \,mK\,s$^{1/2}$\fi}
\def\muKs{\ifmmode \,\mu$K\,s$^{1/2}\else \,$\mu$K\,s$^{1/2}$\fi}
\def\muKRJs{\ifmmode \,\mu$K$_{\rm RJ}$\,s$^{1/2}\else \,$\mu$K$_{\rm RJ}$\,s$^{1/2}$\fi}
\def\muKHz{\ifmmode \,\mu$K\,Hz$^{-1/2}\else \,$\mu$K\,Hz$^{-1/2}$\fi}
\def\MJysr{\ifmmode \,$MJy\,sr\mo$\else \,MJy\,sr\mo\fi}
\def\MJysrmK{\ifmmode \,$MJy\,sr\mo$\,mK$_{\rm CMB}\mo\else \,MJy\,sr\mo\,mK$_{\rm CMB}\mo$\fi}
\def\microns{\ifmmode \,\mu$m$\else \,$\mu$m\fi}

\def\muK{\ifmmode \,\mu$K$\else \,$\mu$\hbox{K}\fi}
\def\microK{\ifmmode \,\mu$K$\else \,$\mu$\hbox{K}\fi}
\def\muW{\ifmmode \,\mu$W$\else \,$\mu$\hbox{W}\fi}
\def\kms{\ifmmode $\,km\,s$^{-1}\else \,km\,s$^{-1}$\fi}
\def\kmsMpc{\ifmmode $\,\kms\,Mpc\mo$\else \,\kms\,Mpc\mo\fi}
\providecommand{\sorthelp}[1]{}

\usepackage{color}
\usepackage{soul}
\usepackage{xcolor}
\usepackage{enumitem}
\usepackage[switch]{lineno}

\def\fsky{\ifmmode f_{sky}\else $f_{sky}$\fi}
\def\microK{\ifmmode \,\mu$K$\else \,$\mu$\hbox{K}\fi}   
\def\microKCMB{\ifmmode \,\mu$K$_{\rm CMB}\else \,$\mu$\hbox{K$_{\rm CMB}$}\fi}
\def\CMBtot{\ifmmode \textit{CMB}_{tot}\else $\textit{CMB}_{tot}$\fi}
\def\deltabetadust{\Delta\textit{Dust}_1}
\def\deltabetadeuxdust{\Delta\textit{Dust}_2}

\def\Nside{\ensuremath{N_{\mathrm{side}}}}

\newcommand{\citedpc}{\citetalias{planck2016-l03}}
\defcitealias{planck2016-l03}{HFI-Legacy}

\newcommand{\sroll}{{\tt SRoll}}
\newcommand{\srolltwo}{{\tt SRoll2}}
\newcommand{\srolltwotwo}{{\tt SRoll2.2}}
\newcommand{\Bcsep}{{\tt Bcsep}}

\begin{document}

\title{\vglue -10mm Improved large scales interstellar dust foreground model and CMB solar dipole measurement}

\author{\small
J.-M. Delouis\inst{1}~\thanks{Corresponding author: J.-M.~Delouis, jean.marc.delouis@ifremer.fr}
\and
J.-L.~Puget\inst{2,3}
\and
L.~Vibert\inst{2}
}

\institute{\small
Laboratoire d'Oc{\'e}anographie Physique et Spatiale (LOPS), Univ. Brest, CNRS, Ifremer, IRD, Brest, France\goodbreak
\and
Institut d'Astrophysique Spatiale, CNRS, Univ. Paris-Sud, Universit\'{e} Paris-Saclay, B\^{a}t. 121, 91405 Orsay cedex, France\goodbreak
\and
Ecole Normale Sup\'{e}rieure, Sorbonne Universit\'{e}, Observatoire de Paris, Universit\'{e} PSL, \'{E}cole normale sup\'{e}rieure, CNRS, Paris, France
}

\date{\vglue -1.5mm \today \vglue -5mm}
\abstract{\vglue -3mm The Cosmic Microwave Background anisotropies are difficult to measure at large angular scales. In this paper, we present a new analysis of the \Planck\ High Frequency Instrument data that brings the cosmological part and its major foreground signal close to the detector noise. The solar dipole signal, induced by the motion of the solar system with respect to the CMB, is a very efficient tool to calibrate a detector or a set of detectors with high accuracy. In this work, the solar dipole signal is used to extract corrections of the frequency maps offsets reducing significantly uncertainties. The solar dipole parameters are refined together with the improvement of the high frequency foregrounds, and of the CMB large scales cosmological anisotropies. The stability of the solar dipole parameters is a powerful way to control the galactic foregrounds removal in the component separation process. It is used to build a model for Spectral Energy Distribution spatial variations of the interstellar dust emission. The knowledge of these variations will help future CMB analyses in intensity, and also in polarization to measure faint signal related to the optical reionization depth and the tensor-to-scalar ratio of the primordial anisotropies. The results of this work are: improved solar dipole parameters, a new interstellar dust model, and a large scale cosmological anisotropies map.} 
 \keywords{Surveys - Cosmology: cosmic microwave background - Cosmology: diffuse radiation - Methods: data analysis - ISM: dust}
\authorrunning{J.-M.~Delouis}
\titlerunning{CMB solar dipole and dust foreground model}
\maketitle

\section{Introduction}
\label{sec:introduction}

Measuring accurately the diffuse extragalactic emissions at low frequencies on the nearly full sky is difficult, especially at large angular scales. It indeed implies being able to separate the atmospheric and the telescope emissions from the astronomical ones and, among those, to separate the extragalactic emissions from the galactic diffuse background ones. The Cosmic Microwave Background (CMB) is the truly diffuse background bringing key cosmological information. At the the peak emission of the CMB frequencies, it imposes to go to space, build cryogenic telescopes and instruments, and find compromises between constraints of the size of telescopes giving a limited angular resolution, and the necessity to cover the whole sky.

In the last 25 years, there has been three generations of CMB space missions COBE, WMAP and \Planck, the next one not providing results before a decade. The Planck High Frequency Instrument (HFI) observed the sky (at 100, 143, 217, 353, 545, and 857\,GHz) with 50 bolometers cooled at 100\,mK. The present and future observations from the ground use large detector arrays, but strongly affected by the atmospheric emission above 200\,GHz. Future space missions will allow to map the CMB and the galactic foregrounds at higher frequencies. Progress of experiments on the ground will certainly outperform \Planck\ capabilities on the CMB itself up to 150\,GHz but not on the high frequency galactic foregrounds, dominating signal at higher frequency.

The large scale polarized $B$ modes is a major goal for cosmology which requires very high sensitivity at CMB frequencies (through more detectors and better control on polarized systematic effects). It also requires better measurements of the low frequency foregrounds with large ground-based surveys, and of high frequency foregrounds hard to achieve from the ground. It is thus a key objective to squeeze out of the HFI data the information on the large scale high frequencies foregrounds. This is the goal of this work.

Although HFI achieved the remarkable result of being cosmic background photon noise limited at the CMB peak frequency at the intermediate and small angular scales (a few degrees to 5\arcmin), at very large angular scales ($\ell<15$), the control of instrument systematics and foreground residuals took a long time to reach the needed level to measure of the cosmological parameter, the reionization optical depth, with a signal to noise ratio of 10 \citep{2020A&A...635A..99P}. The noise and systematic effects residuals at the largest angular scales are nevertheless still far from the detector white noise level needed to accurately remove the high frequency foregrounds. Removing those from the data of a differential detector requires taking advantage of the solar system kinetic dipole. The solar kinetic dipole, common to all detectors and frequencies, is the strongest measurable differential signal at large scales. Improving its determination is thus a key issue throughout this paper. Importantly, the solar kinetic dipole is extracted from various splits of the HFI data (frequencies, galactic masks, component separation models), which constrains an high frequency galactic foreground model. This provides a tool for intercalibration of different instruments and missions as well as constraining a better model of the high frequency foreground systematics at large scales. These improvements, not achieved in the \Planck\ Legacy release, are introduced in this work where we develop a new interstellar dust model (the dominant foregrounds for the CMB above 100\,GHz), self-consistent with a better CMB anisotropies and the associated better solar dipole measurement.

The paper is organized as follow: Sect.~\ref{sec:context} describes the context and goals of the paper; Sect.~\ref{sec:skymodel} describes the algorithm to extract coherently the parameters and templates maps of the foreground, the map zero levels, and presents the resulting dust model; Sect.~\ref{sec:SolarDip_det} build a self consistent CMB anisotropies map and solves for solar dipole parameters.

\section{Context and definitions}
\label{sec:context}
 
The relative motion between the \Planck\ spacecraft and the CMB rest frame generates, a dipolar modulation of the CMB intensity field. The spacecraft motion with respect to the cosmological background is decomposed in two components. 

The first component, referred to as the ``orbital dipole'', is caused by the relative velocity of the spacecraft at the L2 Lagrange point following the earth motion with respect to the solar system rest frame. It is known with a high accuracy at any time. This orbital dipole of order 30\,km\,s$^{-1}$ does not project on the sky maps. It is used to calibrate the absolute photometric response of each detector with a high signal to detector noise of order $10^{-5}$ at the 100 and 143\,GHz CMB dominated channels, and $2\times10^{-4}$ at 353\,GHz \citep[see][hereafter \citedpc]{planck2016-l03}.

The second component, referred to as the ``solar dipole'', is the consequence of the relative velocity of the solar system with respect to the cosmological rest frame. Although it is an order of magnitude larger than the orbital dipole, it cannot be determined from extragalactic astronomy, and thus cannot help to calibrate the absolute response of the detectors. The dipole signal associated with this second component projects onto the sky maps with the same Spectral Energy Distribution (SED) as the CMB cosmological anisotropies. As such, it is not a cosmological effect, but is a systematic effect in the maps extracted by component separation methods on the basis of their SED. It is, nevertheless, not separable from the intrinsic dipolar term of the CMB primordial anisotropies. Using a $\Lambda\-CDM$ model, the CMB anisotropies measurement leads to an expected dipole amplitude of the order of 30\microK. This induces an uncertainty of the order of one percent on the kinetic solar system velocity. This dipolar term is thus included in the definition of the solar dipole, imposing a null dipolar term for the total CMB anisotropies maps.

The solar system velocity measurement is independent of the CMB monopole temperature, known with an accuracy of $3\times10^{-4}$ \citep{fixsen2009}. This introduces additional uncertainty when expressing the intensity amplitude of the solar dipole in $\delta T_{\rm{CMB}}$ anisotropies. Nevertheless, for convenience, all intensity contributions to the dipole amplitudes are expressed in CMB temperature. As such, the conversion between velocity and \microKCMB\ depends directly on the CMB temperature, obtained by fitting a Planck function in the COBE-FIRAS data, which has been refined using a combination of the solar dipole direction from WMAP and the FIRAS full spectrum measurements \citep{fixsen2009}. In this regard, using the solar dipole to estimate the CMB monopole temperature makes to conversion from velocity to \microKCMB\ slightly dependent of the dipole. However, the difference induced by a new dipole determination leads to a negligible correction.

The amplitude of the solar dipole signal is proportional to the response of each detector to a CMB intensity signal, and thus, very useful to test CMB relative photometric calibration of different CMB detectors. Its amplitude of order 3\,000\microKCMB\ is very large with respect to the sensitivity of the 143\,GHz map (amplitude of the noise dipolar $<10^{-2}$\microK). This $3\times10^5$ ratio has been used to measure the relative CMB photometric calibration on single detector maps. The consistency of the relative CMB photometric calibration on single detector maps within a frequency has a noise limited accuracy of $10^{-5}$. The inter-calibration between HFI frequency maps is at the $4\times10^{-4}$ level from 100 to 353\,GHz (\citedpc). This work uses these sensitivities, and exploit the fact that all detectors measure the same CMB solar dipole (called hereafter ``the consistency argument'') to control the residuals from foregrounds on the largest angular scales.

The CMB map is obtained from the frequency maps using its properties: Planck function SED and Gaussian statistics. The extraction of solar dipole amplitudes measured at various frequencies and on different sky fraction is the most powerful tool to detect the foreground residuals. This was used for example for the HFI data showing residuals behaving with frequency like a galactic dust component as can be seen in Tables 7 and 8 of the \citedpc\ are obviously due to the dust foreground removal, and showed that introducing large scale dipolar and quadrupolar terms to account for the large scale spectral energy distribution (SED) variation reduces very much these drifts for the highest frequencies and sky fraction used to fit the solar dipole.
 
Here, the solar dipole is extracted with an improved version \srolltwo\ \citep{2019A&A...629A..38D} of the \sroll\ algorithm used to produce the \Planck\ Legacy HFI maps. This upgraded algorithm, named \srolltwotwo, introduces improvements: 
\begin{itemize}
\item Use of 143, 217, 353, 545, and 857\,GHz non-polarized Spider Web bolometers (SWB) data, in addition to Polarized Spider bolometers (PSB) data. This allows for reduced systematic effects for intensity only maps, while avoiding being limited by signal to noise ratios;
\item Removal of polarized parts of the signals due to the small parasitic sensitivity to polarization of SWBs detectors;
\item Improved correction of the Analog-to-Digital Converters non-linearities, bringing the associated residuals below detector noise levels, even at the largest angular scales;
\item Computation of single detector maps removing polarization part in the signal;
\item Use of the full resolution for the bandpass leakage correction where the signal is very strong, keeping the low resolution of the foreground templates on the rest of the maps to avoid introducing extra noise;
\item Removal the Cosmic Infrared Background (CIB) monopole \citep{1999A&A...344..322L} and use CMB maps with zero monopole;
\item Setting the initial zero level of the intensity maps to the zero level of the 21-cm emission of the \ion{H}{i} galactic gas foreground.
\end{itemize}

\srolltwotwo\ builds two types of single detectors intensity maps\footnote{Those maps are available at \href{http://sroll20.ias.u-psud.fr}{http://sroll20.ias.u-psud.fr}} where: i) the correction of all foreground signal bandpass mismatch between detectors is applied, ii) no correction of the foreground signal bandpass mismatch is applied. Those latest maps, smoothed at a common 1\deg, are used in this work.

\section{Foreground sky model and zero level offsets}
\label{sec:skymodel}

The solar dipole amplitudes of \cite{planck2014-a10} Table 2, are drifting away from the amplitude measured at 100 and 143\,GHz both for higher frequencies (217 to 545\,GHz), and when increasing the sky fraction \fsky\ used to fit the dipole. The increase of the amplitude errors with frequency follows a typical galactic dust SED. Such variations were expected from previous work on the COBE-DIRBE data \citep{1998A&A...333..709L}. \citedpc\ shows that this is mostly cured by introducing, in the model, large scale variations of the dust SED. This result is obtained using the stability of the solar dipole direction, taking the 100\,GHz one as reference. The dipole direction (especially the longitude) drifts by 0.1\deg but the parameters for the 545\,GHz still exhibits large shift in amplitude and direction. 

These results indicate that the residuals from the dust emission removal are a major concern when extracting the solar dipole above 100\,GHz in two ways:
\begin{itemize}
\item first, indirectly by the dust residuals along the galactic ridge (central parts versus anticentre of the galactic disc). This forces the introduction of a galactic ridge mask $M_g$ where the the CMB anisotropies cannot be removed accurately enough. This induces an unavoidable error when the dipolar term of the cosmological anisotropies is set to zero on the full sky (see Sect.~\ref{sec:dipolerem});
\item second, the large scale galactic emission in the part of the sky where the solar dipole is fitted outside the galactic ridge mask affects the residuals of the component separation process. We thus introduce, in our foreground dust emission sky model, terms improving the modeling of the large scales SED variations.
\end{itemize}

The CMB anisotropies component of the maps, as well as the noise and systematic effects, are defined with a null monopole term. The HFI frequency maps do not have an absolute zero setting calibration. Any error in the zero level changes the contrast between the high and low galactic latitude foregrounds, thus changes the foreground maps, especially where the intensity is very low at high latitude. The large scale residuals of dust emission depends strongly on the accuracy of these zero levels, and we introduce consistency constraints from the solar dipole to refine the accuracy of the foreground model.

To start with, the initial setting of the zero level of the \Planck\ Collaboration HFI frequency maps use the method in which the extrapolated zero intensity emission of the galactic foregrounds gas correspond to the extrapolated zero column density of gas \citep{planck2013-p03f}. The first step is performed by regressing the 857\,GHz, fully dominated by the dust emission intensity, to the neutral hydrogen interstellar gas column density. Then the other frequencies are regressed to the 857\,GHz emission brightness. This sets the initial zero levels of the frequency maps good enough accuracy as long as one does not require highly accurate intensity on very large scales.

To improve on these results, and especially on the dust template map zero level, a more accurate method is developed here\footnote{The full schematics of the algorithm is available at \href{http://sroll20.ias.u-psud.fr}{http://sroll20.ias.u-psud.fr}}.

\subsection{Dust model definition}
\label{sec:Dustmodeldef}

We decompose the intensity sky model as:
\begin{equation} 
\label{eq:datamodel}
\textit{sky model}\sim \textit{CMB}_{tot} + \textit{FreeFree} + \textit{Synchrotron} + \textit{CO} +dust
\end{equation}
In Eq.~\ref{eq:datamodel}, there is only one extragalactic component, the CMB anisotropies. The CIB large scales ($\ell<100$) anisotropies are negligible with respect to the galactic dust component, and are included in it. \CMBtot\ is the total CMB anisotropies map containing the solar dipole and the primary cosmological anisotropies. The \CMBtot\ intensity is described by a single map of $\delta T_{\textit{CMB}}$ at all frequencies.

The three next terms of Eq.~\ref{eq:datamodel} are the low frequency foregrounds: free-free, synchrotron emission, and CO emission lines. In intensity, the free-free and synchrotron emissions are respectively one, and two orders of magnitude smaller than the CMB anisotropies at 100\,GHz. The free-free emission, which is only one order magnitude lower than the dust at 100\,GHz, benefits from a stable and accurately known SED. Furthermore, the synchrotron emission is lower than other emissions by orders of magnitude at higher frequencies. Thus free-free and synchroron emissions are removed in an open loop, using \Planck\ Legacy results \citep{planck2014-a12}. In the HFI data, the CO lines emission are derived using only the large differences of response between detectors within a frequency band containing a CO line and the \Planck\ Legacy CO template maps are used. Nevertheless, their amplitude are updated, consistently with the new dust model.

The last term of Eq.~\ref{eq:datamodel}, the interstellar dust emission, is the dominant foreground at frequencies above 100\,GHz. It is described by a new model:
 \begin{equation} 
\label{eq:dustmodel}
\textit{dust} = f(\nu) Dust + \frac{\partial f(\nu)}{{\partial \beta}} \deltabetadust + \frac{\partial^2 f(\nu)}{\partial \beta^2} \deltabetadeuxdust \quad ,
\end{equation}
where $f(\nu)=\textit{Planck function}(\nu,T) \times \left(\frac{\nu}{\nu_0}\right)^{\beta}$, $\nu_0$ beeing the reference frequency of the $\textit{Dust}$ map.

The temperature $T$ and the index $\beta$ are very correlated when fitted in the range of frequencies used in this work, thus the temperature is taken as constant, and the SED variations are only described by the variations of the $\beta$ parameter. Given the available number of degrees of freedom (5 frequencies), this has been shown acceptable in the HFI frequency range from previous work using a broader frequency range from the absolute intensity maps from COBE. The single temperature adopted 18\,K has been shown to give good SED fits for the \ion{H}{i} dominant phase of the interstellar medium but also for the subdominant \ion{H}{+}, that show very similar SED \citep{1998A&A...333..709L}.

The novelty of this dust model is the use of an analytical description of the averaged SED, and the introduction of two additive correction modeling the effects of the SED spatial variations.

The first term of Eq.~\ref{eq:dustmodel}, $Dust$, is the initial dust map (see further), propagated from the frequency $\nu_0$ to each single detector map with a frequency $\nu$. Furthermore, the frequency is taken at the effective frequency that takes into account the difference of the response of the detector to a dust SED with the response to the CMB orbital dipole signal used to calibrate the detector with high accuracy in K$_{\text{CMB}}$.

The two other terms of Eq.~\ref{eq:dustmodel} describe the spatially variable part of the SED, absorbing all differences to the spatially constant SED first term. The analytical description of the SED $f(\nu)$ allows a description of these variations as an expansion in term of its first and second derivatives with respect to $\beta$. The model is not strictly described by the analytical expansion of the spectral energy distribution function $f(\nu)$ because the second spatial template $\deltabetadeuxdust$ is independent from the first one, $\deltabetadust$, which describes the first order expansion term. These two independent full sky maps with a null average have a resolution of order 1\deg ({\tt HEALPix} \citep{gorski2005} \Nside=128).

The initial dust map $Dust$ is built from a map at high enough frequency ($\nu_0$=545\,GHz) that it only needs to be cleaned from the CMB, the CO, and the free-free emissions in an open loop. The CMB large scales cosmological anisotropies map, taken from the \Planck\ Legacy release, once propagated to $\nu_0$, are then three orders of magnitude smaller than the dust emission. The effects of the difference between the initial CMB used at this stage with the CMB map obtained ultimately in this work, is thus negligible.

From Eq.~\ref{eq:dustmodel}, we obtain the dust map model $\mathcal{D} (\nu_b)$ for a single detector (bolometer $b$) map:
\begin{equation} 
\label{eq:finaldustmodel}
\mathcal{D}(\nu_b)=\frac{f(\nu_b)}{f(\nu_0)} \left(\!Dust_{\nu_0}+\ln(\frac{\nu_b}{\nu_0}) \deltabetadust\!+ \left( \ln (\frac{\nu_b}{\nu_0}) \right)^2\! \deltabetadeuxdust \right) .
\end{equation}
This equation is valid for maps expressed in spectral energy density. The 545 and 857\,GHz maps are calibrated on planets and are dealt with in spectral energy density expressed in \MJysr. Nevertheless, 100-353\,GHz detectors, calibrated on the orbital dipole of the CMB in $\delta T_{CMB}$, are expressed in \microKCMB, and must be converted to spectral energy density. The lower frequencies CMB maps, offsets and solar dipole amplitudes are finally be converted back to \microKCMB\ for the convenience of comparison with similar maps in the literature.

We know that SED variations exist at intermediate to large scales due to the gradients of i) stellar radiation field intensity changing the dust temperature, ii) the relative fraction on each line of sight of the molecular, atomic and ionized gas fraction which are known to have slightly different dust properties. The two SED variation components are thus defined by maps smoothed at 1\deg.

\subsection{Extraction of the CO emission lines, dust model parameters and templates}
\label{sec:SEDmap}

The \srolltwotwo\ single detector intensity maps are built after removing the polarized signal and the bandpass mismatch coefficients have been computed using template dust maps different from the ones build in the present work. We thus use the set of maps in which the color correction has not been corrected to be able to extract band pass mismatch coefficients coherent with the dust model using an iterative process.

We name $I_b$ the single detector intensity map built from the bolometer $b$ data. The effective frequency $\nu_b$, computed from the bandpass ground measurements, is used in the first iteration of the processing, and is corrected at each iteration when regressing the single detector map against the dust initial template $Dust_{\nu_0}$.

The dust emission is the only broad band foreground component not fully described by its SED, but modeled empirically for the frequencies $\nu\ge100$\,GHz as a map with an average SED, completed with two additive correction maps describing the spatially variable SED. These three maps are extracted in this work, using other constraints than previous component separation ones as discussed above (e.g. consistency of the solar dipole parameters). The CO and dust foregrounds being partially correlated, the CO foreground map is built using a CO template map from the \Planck\ Legacy results \citep{planck2014-a12} but whose coefficients are extracted coherently with our new dust model.

Finally, the data contains noise and monopole offsets, not directly measurable by the HFI differential detectors, and which are to be solved for, using consistency arguments and redundancies.

Equation~\ref{eq:residualmap} defines the residual $R_b$ which only contains the \CMBtot\ and the foreground components to be modeled. It is computed for each single detector map $I_b$ from which are removed, in an open loop, the low frequency foreground maps taken from the \Planck\ legacy maps as described above.
\begin{equation} 
\label{eq:residualmap}
R_b=I_b-\textit{FreeFree}_b-\textit{Synchrotron}_b - O_{HI} \quad .
\end{equation}
The initial offset $O_{HI}$ applied to the single detector map sets the zero level of the single detector intensity maps $I_b$, following \cite{planck2013-p03f} that sets this zero level to the extrapolation to zero column density of HI 21-cm emission of interstellar gas at high latitudes. This is a first approximation of the absolute zero level of the dust foregrounds. The addition of the ionized component traced by H$_\alpha$ emission changes the zero level by 3\% within the uncertainties, and the molecular component is negligible at high galactic latitudes. The offsets are further refined in this work with smaller uncertainties, and the final results should be consistent with the initial offsets. The offset values for detectors, either within the same frequency band or with different frequencies, change the contrast of the large scales at high latitude but are constrained by the solar dipole signal which is a common component to all bolometers.

For the 100 to 353\,GHz detectors, the single detector map offset $O_{b} $ is adjusted in three steps:
\begin{enumerate} 
\item The first step uses all pair differences that gives offsets with respect to a zero average offset of all these bolometers. 
\item The second step adjusts the average offset by using the fact that the solar dipole vector seen by all the detectors is the same.
\item The third step is the determination of the 545\,GHz frequency map offset, performed by minimizing its projected effect by the dust emission SED from 545\,GHz to the lower frequencies offsets. The initial dust template is extracted from the 545\,GHz frequency map, which is corrected at the end of this step. This provides a new dust template $Dust_{\nu_0}$, used at the next iteration.
\end{enumerate}

The residual $R_b$ also contains the not-recoverable noise and systematic effects residuals $N_b$ (including the very small residuals following the low frequency foregrounds removal).

In order to improve the solar dipole measurement, an improved model of the two main partially correlated high frequency foregrounds components is built extracting simultaneously the CO lines emission intensity coefficients $\gamma_b$ and the effective frequencies $\nu_b$ and templates maps $\deltabetadust$ and $\deltabetadeuxdust$ describing the new thermal emission dust model. The constraint used is, for each sky pixel seen by all pairs of bolometers $b1$ and $b2$, to minimize the difference $\Delta R_{b1,b2}= R_{b1} - R_{b2}$, expressed in \microKCMB. The CMB$_{tot}$ term cancels and:
\begin{equation} 
\label{eq:diffbolo}
\Delta R_{b1,b2} =
\mathcal{D}(\nu_{b1})-\mathcal{D}(\nu_{b2}) +\!\left(\gamma_{b1}\!-\!\gamma_{b2}\right)\widetilde{CO}\!+ O_{b1}\!-\! O_{b2} + N_{b1}\!-\!N_{b2}
\end{equation}
where $\mathcal{D}(\nu_{b})$ is the dust emission (Eq.~\ref{eq:finaldustmodel}). The initial dust map $\mathcal{D}(\nu_0)$ is computed from the \srolltwotwo\ 545\,GHz map, after subtracting in an open loop CO, free-free and CMB. For bolometer $b$, the specific projection coefficient $f(\nu_b)/f(\nu_0)$ from the template dust map computed from 545\,GHz bandpass map, account for the detector response to the dust SED, through the effective frequency $\nu_b$. These effective frequencies, initially set to the values computed from the ground measurements of the bandpass \citep{planck2013-p03d}, are adjusted simultaneously with the $\gamma_b$ coefficients. 

The CO \Planck\ Legacy maps are used as a template to extract a single effective set of coefficients the CO lines response $\gamma_b$ at 100, 217, 353\,GHz. The fact that there is no CO line in the 143\,GHz frequency band is used to regularize the solution, requiring that the detectors have a negligible response correlated with the $\widetilde{\textit{CO}}$ template. Weaker interstellar molecular lines nevertheless exist in the 143\,GHz frequency band, but this affects the CO coefficients by an error of 1.1\%, inducing an error on the solar dipole amplitude of 0.06\microK\ and $10^{-4}$ degree in direction, thus negligible. 

The offsets $O_b$ are strongly degenerate with the dust large scale SED variation, thus are also solved for simultaneously using the same minimization of the differences $\Delta R_{b1,b2}$ used to compute the dust large scale SED variation. The quantity to minimize, based on the internal stability of the frequency maps, is, for each pixel $p$:
\begin{multline}
\label{eq:chi2dust}
\chi_{hf}^2=\sum_{(b1, b2)} \sum_{p}\left( \Delta R_{b1,b2,p} - \mathcal{D}(\nu_{b1})+\mathcal{D}(\nu_{b2}) \right. \\
\left. \!-\!\left(\gamma_{b1}\!-\!\gamma_{b2}\right)\widetilde{CO}_p\!-\!O_{b1}\!+\!O_{b2}\!\right)\!^2 \quad .
\end{multline}
Conditions on the mean parameter values between all single detector are needed to minimize Eq.~\ref{eq:chi2dust} that is based on map differences. This is possible thanks to several closure conditions.
\begin{itemize} 
\item For $O_b$ , $\deltabetadust$, and $\deltabetadeuxdust$, we impose the mean values to be null.
\item For $\gamma_{b}$, we force the mean values to be null at 143\,GHz.
\item The level of residual dust emission in the mean map over all detectors after removing the dust and the CO model, and an initial CMB anisotropies map, is related to a miss estimation of the mean $f(\nu_{b})$. Setting this level to $0$, provides a closure condition for all $\nu_{b}$. At the first iteration, we use the CMB {\tt SMICA} map. 
\end{itemize}

From this minimization, we obtain:
\begin{itemize} 
\item the correction to the response to a dust SED for bolometers calibrated on the CMB orbital dipole as an effective frequency $\nu_b$,
\item the additive SED spatial variation described by the $\deltabetadust$ and $\deltabetadeuxdust$ maps,
\item the CO effective coefficients $\gamma_b$,
\item the relative offsets $O_b$ of the single detector maps. 
\end{itemize}

Furthermore, a global 100-353\,GHz offset $\bar{O}$ must be introduced to adjust the offsets of the 100-353\,GHz maps with respect to the unknown 545\,GHz map offset. Moreover, considering the large uncertainty on the 545\,GHz map offset, a correction has also to be introduced. This will correct the large scales initial dust map template taken from the 545\,GHz frequency map. This is performed by adding a constraint on the solar dipoles built for each detector (Sect.~\ref{sec:Dustoffset}).

\subsection{Determination of the frequency map offsets}
\label{sec:Dustoffset}

The dust SED spatial variations are strongly degenerate with the average zero levels of the frequency maps. Equation~\ref{eq:chi2dust} minimization only constrains the relative adjustments of these offsets, and do not affect the initial band average zero levels of the frequency maps which have rather large uncertainties with respect to the relative ones, especially for the 353 and 545\,GHz frequency band.

A global (100 to 353\,GHz) offset $\bar{O}$ is introduced in the algorithm. It is determined thanks to the fact that the spatial distortions of the dust foreground initial dust map uncertainties, biases the solar dipole vector directions. Thus, the minimization of the dispersion of the single detector CMB solar dipole vector over the 43 (100 to 353\,GHz) HFI bolometers allows to compute $\bar{O}$.

To remove the solar dipole, the cosmological CMB anisotropies map must be removed from the \CMBtot\ map. Nevertheless this CMB anisotropies map, extracted by component separation methods, is not reliable in the narrow galactic ridge that needs to be masked and filled with a constrained realization within a galactic ridge mask $M_g$ before setting the dipole term to zero on the whole sky. The mask $M_g$ is defined as the sky contained between two symmetrical latitudes centered on the galactic plane and characterized by the sky fraction \fsky\ it covers.

The CMB anisotropies full sky map noted $\textit{CMB}_{M_g}$ is the \CMBtot\ map from a component separation masked with $M_g$ in which the dipole term has been set to zero. It depends on the size of the galactic mask $M_g$, and on the procedure used to fill this mask with CMB anisotropies constrained realizations. Before having a \CMBtot\ anisotropies map consistent with the new foreground model, we use, at the first iteration, the \Planck\ Legacy CMB {\tt SMICA} map, estimated reliable for cosmology outside a $5\%$ galactic mask. At the second and following iterations, a CMB large scales anisotropies map, coherent with Sect.~\ref{sec:skymodel}, result can now be built. We remove from all single intensity bolometer maps $R_b$ from 100 to 353\,GHz, the current dust model and CO template, and the adjustments of the offsets obtained in Sect.~\ref{sec:skymodel}. The weighted average over all bolometers provides the new current \CMBtot. 

A new three-components (CMB, dust and CO) separation algorithm, named \Bcsep, is developed using difference between single detector maps in nested iterative loops, also using the consistency of the solar dipole parameters between frequencies. This requires to start with good enough approximations and the convergence of the algorithm is the test of this condition. This new component separation is performed by subtracting the dust or the CMB maps from the frequency residual map $R_b$ and improving iteratively \CMBtot\ and the three maps of the dust model per frequency:
\begin{equation} 
\label{eq:cmb}
\textit{CMB}_{tot}=\left( \sum_b\! \frac{1}{\sigma_b^2}\right)^{-1}\! \sum_b\frac{1}{\sigma_b^2}\left(R_b-\! \mathcal{D}(\nu_{b})-\! \gamma_{b}\widetilde{CO}-\! O_b-\! \bar{O}\right) .
\end{equation}
where $\sigma_{b}^{2}$ is the variance of the cleaned maps for \fsky=0.9 estimated by simulations. 

Equation~\ref{eq:residualcmb} defines the single detector intensity map $R_{b}$ as the residual map $\aleph_{b,M_g}$ after removal of the high frequency foregrounds model, the current CMB anisotropies $\textit{CMB}_{M_g}$, and the relative offsets extracted previously. $\aleph_{b,M_g}$ is thus the map containing only the CMB solar dipole, together with the noise and residuals from the foregrounds removal:
\begin{eqnarray} 
\label{eq:residualcmb}
\aleph_{b,M_g} & = & R_{b} - \textit{CMB}_{M_g} - \mathcal{D}_{b} - \gamma_{b} \widetilde{CO} - O_b \\ \nonumber
&=& \text{A}_{b,M_g, \bar{O}} \sum_{\ell=1, m=0, \pm1} \mathcal{A}_{b,M_g}\mathcal{Y} + \bar{O} + N_{b,M_g} \quad ,
\end{eqnarray}
where
\begin{itemize} 
 \item A$_{b,M_g, \bar{O}}$ is the solar dipole amplitude extracted from $\textit{CMB}_{M_g}$,
\item $\mathcal{A}_{b,M_g,\bar{O}}$ are the three $a_{\ell,m}$ vector components ($a_{1,0}$, $a_{1,1}$ real and imaginary parts) of the solar dipole direction.
\item $\mathcal{Y}$ are the three spherical harmonics Y$_{\ell=1,m}$ component maps of the solar dipole.
\item $\bar{O}$ is the global offset of the detector maps 100-353\,GHz initially set to zero.
\end{itemize} 

To constrain $\bar{O}$, we impose that all detectors show a minimal dispersion of the solar dipole direction vector. Starting with $\bar{O}=0$, we loop on Eqs.~\ref{eq:residualcmb}, \ref{eq:dipolefitnew}, and \ref{eq:chi2new} and solve for the $\bar{O}$ value. The solar dipole parameters for bolometer $b$ are fitted within a set of 10 masks $\mathcal{M}$ taken in the sky region left outside the galactic ridge $M_g$ to smaller and smaller areas identified by their \fsky. 

The minimization Eq.~\ref{eq:dipolefitnew} extracts, for a given detector, a running $\bar{O}$ and mask $M_g$, the dipole amplitude $\text{A}_{b,M_g, \bar{O}}$ and direction vector $\mathcal{A}_{b,\mathcal{M}_p, \bar{O}}$ by minimizing the difference taken over for a mask $\mathcal{M}$ between the residual map $\aleph_{b,p,M_g}$ map and the model solar dipole map plus the current $\bar{O}$:
\begin{equation} 
\label{eq:dipolefitnew}
\chi^2_{b,\mathcal{M}, M_g}\! =\! \sum_p \mathcal{M}_{p}\left(\aleph_{b,p,M_g}\! -\! \text{A}_{b,M_g, \bar{O}}\! \sum_{\substack{\ell=1\\ m=0, \pm1}}\! \mathcal{A}_{b,\mathcal{M}_p, \bar{O}}\mathcal{Y}_p\! -\! \bar{O} \right)^2.
\end{equation}

Then $\bar{O}$ in carried back into Eq.~\ref{eq:residualcmb} if we change $M_g$ or into Eq.~\ref{eq:dipolefitnew} if we change only $\mathcal{M}$ after the optimization of $M_g$, until convergence.

This should not be affected at the first iteration by the choice of the CMB cosmological anisotropies {\tt SMICA} map as it converges to a stable $\bar{O}$ value.
\begin{equation} 
\label{eq:chi2new}
\chi_{b,M_g}^2 = \sum_{b=1}^{N_b}\sum_{\mathcal{M}=1}^{N_{\mathcal{M}}} \left(\mathcal{A}_{b,\mathcal{M}_p, \bar{O}} - \sum_{b=1}^{N_b}\sum_{\mathcal{M}=1}^{N_{\mathcal{M}}}{\frac{\mathcal{A}_{b,\mathcal{M}_p, \bar{O}}}{N_{\mathcal{M}} N_b}}\right)^2 \quad .
\end{equation}

At this stage, the 545\,GHz map initial offset uncertainty still affects all lower frequencies zero levels indirectly through the distortion of the initial dust template and the minimization Eq.~\ref{eq:chi2dust} to Eq.~\ref{eq:chi2new}, and directly by the propagation of the 545\,GHz offset residual to lower frequencies through the SED $f(\nu)$. 

We define the 100 to 353\,GHz offsets $O_{\nu}$ as the average of $(O_b+\bar{O})$ over the frequency band $\nu$. The correction of the initial offset at 545\,GHz, $O_{545}$, which also affects all the $O_{\nu}$ (following the $f(\nu)/f(\nu_0)$ emission law), is the one which minimizes Eq.~\ref{eq:chi2offset}:
\begin{equation} 
\label{eq:chi2offset}
\chi^2 = \sum_{\nu=100}^{353}\frac{1}{{{\sigma}_{HI}}^2} \left( O_{\nu} - \left(\frac{f(\nu)}{f(\nu_0)}     \right)O_{545}\right)^2 \quad ,
\end{equation}
where ${\sigma}_{HI}$ is the rms of the initial zero level determination uncertainties at frequency $\nu$. Thus this correction will affect mainly the 353\,GHz.

The next section (Sect.~\ref{sec:offsetrem}) discusses the coherence of the offset determination.

\subsection{Absolute zero level correction consistency}
\label{sec:offsetrem}

The absolute zero levels of the maps are critical for the knowledge of the foregrounds that need to be modeled with the right zero level. In Sects.~\ref{sec:SEDmap} and \ref{sec:Dustoffset}, we develop a method to constrain these offsets using the solar dipole strong signal that should be seen in all single detector maps with the same direction vector.

Four map offsets are determined in sequence:
\begin{enumerate}
\item The initial setting of the zero levels in the \srolltwotwo\ maps $O_{HI}$ are computed following \cite{planck2013-p03f}. Their uncertainties, although small in the CMB frequencies, affect the dust emission large scales at high latitudes at 353 and 545\,GHz.
\item Equation~\ref{eq:chi2dust} describes the computation of the relative offsets of the 100 to 353\,GHz single bolometer maps, $O_b$. The constraint is the consistency between all pairs of single bolometer maps. To allow the minimization based on differences of maps, the mean of all these relative offsets per single bolometer is set to 0.
\item Section~\ref{sec:Dustoffset} deals with the zero level mean correction over all 100 to 353\,GHz bolometers, $\bar{O}$. It is obtained by minimizing, over all single detectors and masks, the variance of the solar dipole directions using Eq.~\ref{eq:dipolefitnew}. The single bolometer solar dipoles are extracted using Eq.~\ref{eq:chi2new}. Finally, we mesure $\bar{O}=$4.74\microK. The relative corrections are bigger on the lowest frequencies.
\item Finally, Equation~\ref{eq:chi2offset} adjusts the 545\,GHz offset, $O_{545}$, by minimizing the lower frequencies offset corrections affected by the projection of $O_{545}$ following the dust SED $f(\nu)$. This offset is applied to the initial dust map $Dust_{\nu_{545}}$, and the whole process is iterated over.
\end{enumerate}
The results after convergence are reported in Table~\ref{tab:offsetcomputation}.
\begin{table}[h] 
\newdimen\tblskip \tblskip=5pt
\caption{For the HFI frequency bands, $O_{HI}$ gives the initial map offsets based on the zero HI column density extrapolation; $O_b$ gives the relative bolometer offsets averaged over the frequency band; column $+\bar{O}$ gives the offsets after adding the global offset $\bar{O}$; column $Final$ gives the final offsets after removing the 545\,GHz offset. Last column gives the final uncertainty on these zero levels.}
\label{tab:offsetcomputation}
\vskip -6mm
\footnotesize
\setbox\tablebox=\vbox{
\newdimen\digitwidth
\setbox0=\hbox{\rm 0}
\digitwidth=\wd0
\catcode`*=\active
\def*{\kern\digitwidth}
\newdimen\signwidth
\setbox0=\hbox{+}
\signwidth=\wd0
\catcode`!=\active
\def!{\kern\signwidth}
\halign{\tabskip 0pt\hbox to 1.2cm{#\leaderfil}\tabskip 0.2em&\hfil#\hfil\tabskip 1em&\hfil#\hfil\tabskip 0.5em&\hfil#\hfil\tabskip 0.5em&\hfil#\hfil\tabskip 0.5em&\hfil#\hfil\tabskip 0em\cr
\noalign{\doubleline}
\omit&\hfil\sc Planck\hfil& \multispan4\hfil\sc This work\hfil\cr
\noalign{\vskip -5pt}
\omit&\hrulefill& \multispan4\hrulefill\cr
\omit \hfil Frequency\hfil& $O_{HI}$ & $O_b$ & $+\bar{O}$ & $Final$ & rms\cr
\omit \hfil [GHz]\hfil & \hfil[\muK] & \hfil[\muK] & \hfil[\muK] & \hfil [\muK] & \hfil [\muK]\cr
\noalign{\vskip 3pt\hrule\vskip 5pt}
100& -3.5$\pm$3.6& -3.6*  & *1.14 & 1.05  & *1.0 \cr
143& -8.28$\pm$2.8 & -1.08 & *3.67 & 3.49 & *1.1\cr
217& -26.1$\pm$5.8 & -2.87 & *1.87 & 1.24 & *2.4\cr
353& -142.4$\pm$27 & 6.26 & 11.0* & 6.17 & 12.2\cr
\noalign{\vskip 3pt\hrule\vskip 5pt}
\omit \hfil \hfil&\hfil[\MJysr]\hfil &\hfil &\hfil &\hfil[\MJysr]\hfil &\hfil [\MJysr]\hfil \cr
545& -0.884$\pm$0.012 &          &          & \phantom{-}0.0048   & 0.0023 \cr
857& -0.534$\pm$0.16* &          &          & -0.124*   & 0.04** \cr
\noalign{\vskip 3pt\hrule\vskip 5pt}}}
\endPlancktablewide
\end{table}
The data zero level offset corrections, averaged per HFI frequency band, are given as cumulative offset values obtained after each correction of the algorithm refining the initial offset $O_{HI}$. The convergence of the iterative process shows the stability of the algorithm which leads to a minimal dispersion of the solar dipole vector direction for the single detector CMB anisotropies maps in the frequency bands 100-353\,GHz.

The final absolute zero level uncertainties are computed using a set of ten simulations for the four lower frequencies. Those simulations have representative noise and systematic effects but no attempt is done to quantify the uncertainties due to the dust model by simulations which have no reliable statistical model. These uncertainties at 100, 143, and 217\,GHz are reflected in the dispersion of the single bolometers at each frequency. Thus, these errors are probably dominant over the instrumental systematic effect dispersion. The uncertainty at 353\,GHz is about one third of the dispersion between bolometers which indicate that the zero level correction reflects the dust foreground effects. The uncertainty at 545\,GHz is computed using a Monte Carlo approach based on uncertainties at the lower frequencies. The $O_{545}$ correction induces a marginally significant correction at 353\,GHz only. These corrections steps are weakly correlated, and the convergence confirms that the algorithm steps are efficient to improve the offsets. 

\begin{figure}[ht!] 
\includegraphics[width=\columnwidth]{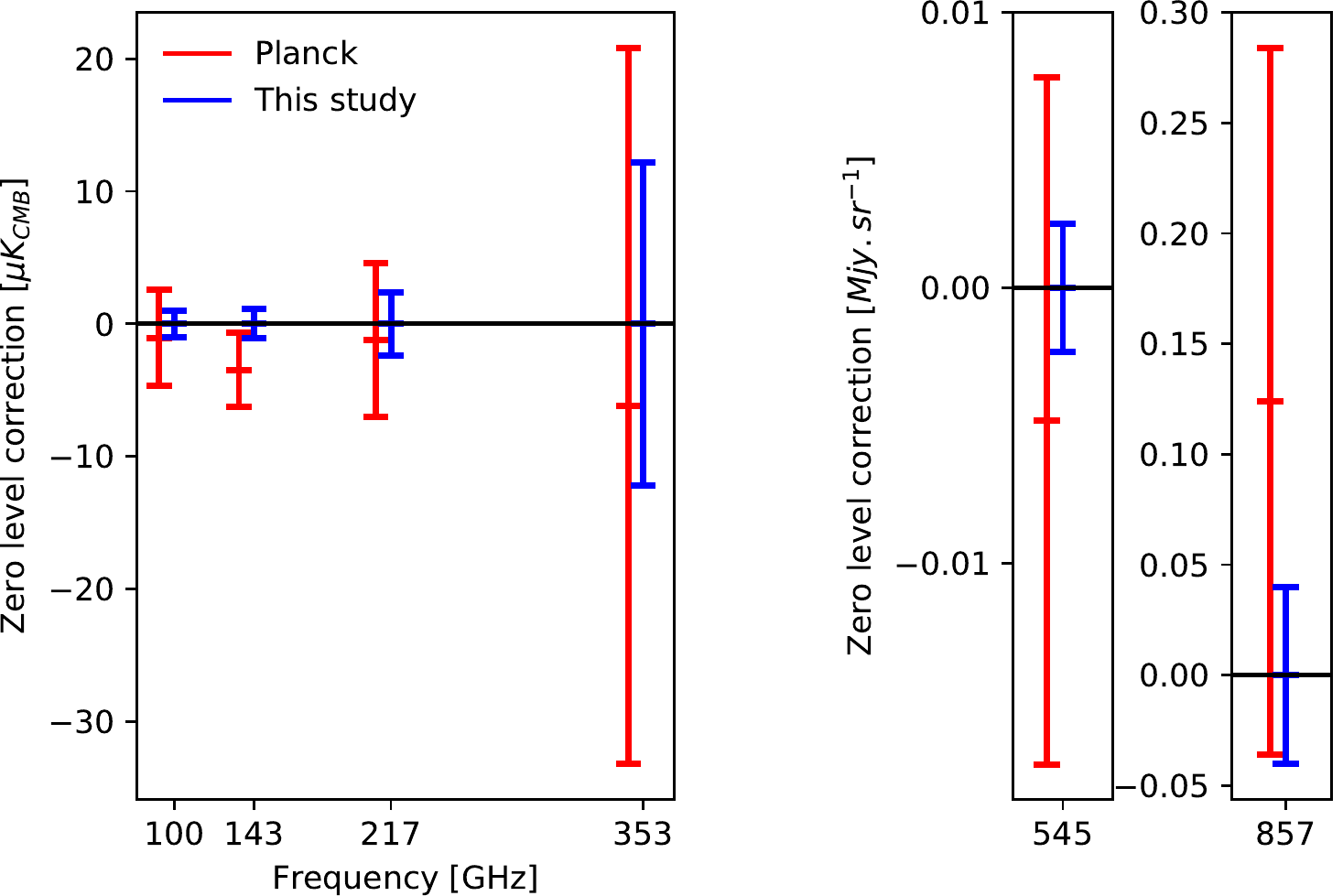} 
\caption{Zero levels of the maps corrections, in red the initial one, and in blue the final cumulated offsets corrections taken as reference, with their uncertainties.}
\label{fig:ZERO_LEVEL} 
\end{figure}

Figure~\ref{fig:ZERO_LEVEL} displays the map offset results, the initial ones from \Planck\ in red, and the final ones as a function of the frequency bands.
The offset corrections from this work, in blue, are consistent with the initial frequency map offsets but with much better accuracies by factors 2.5 to 5.

\subsection{Dust model results}
\label{sec:Dustmodel}

\begin{figure}[htbp!] 
\includegraphics[width=\columnwidth]{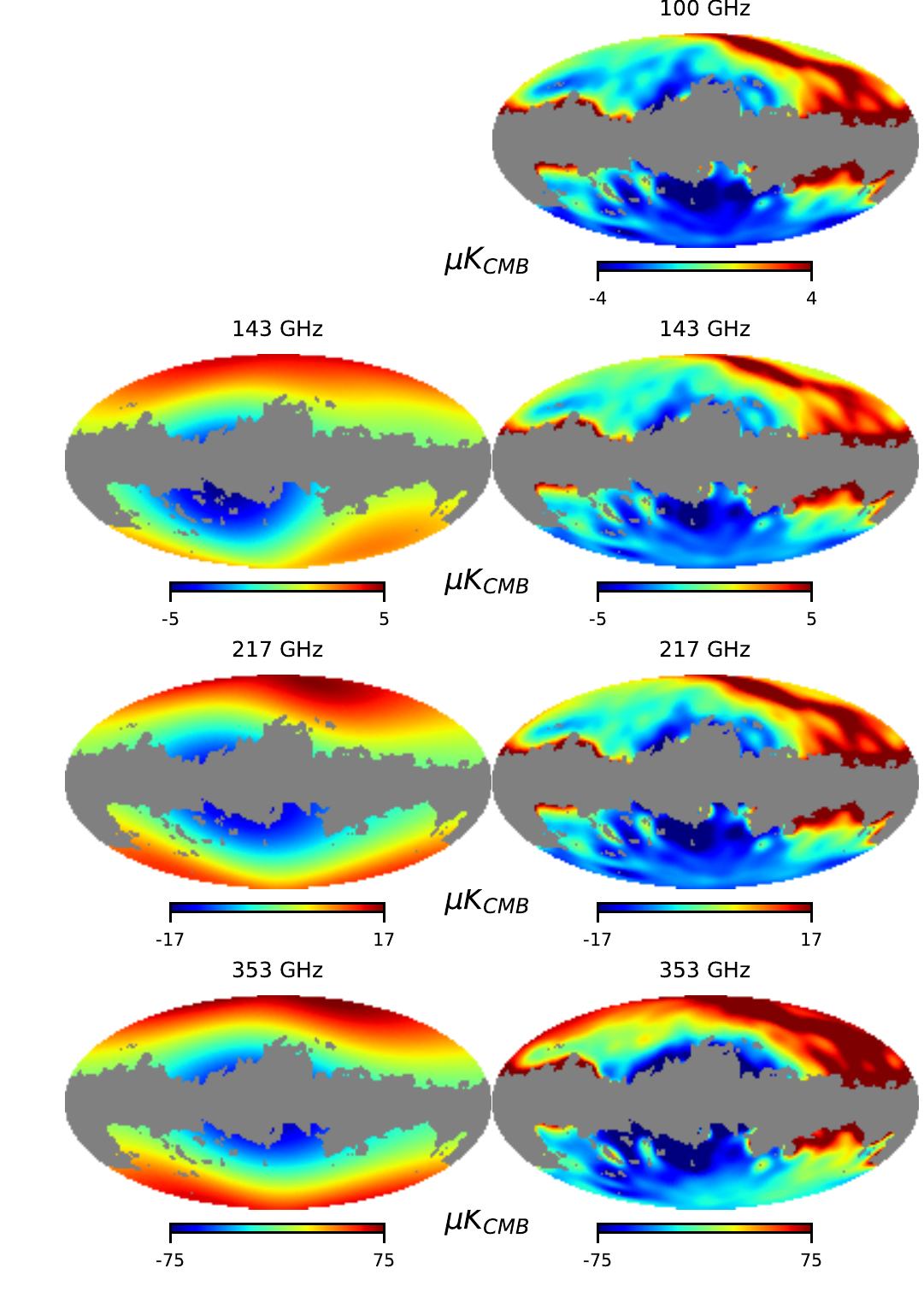}
\caption{Right column shows the additive correction maps from this work. The equivalent correction maps from \citedpc, which only contains dipole and quadrupole terms, are shown in the left column. The 100\,GHz map is not displayed as it was used as the reference for the dipole direction.}
\label{fig:dipresidual} 
\end{figure}

Figure~\ref{fig:dipresidual} presents these additive correction maps (as {\tt Healpix} maps at \Nside=128) to take into account the spatial variation of the dust emission SED. These spatial variation maps, in the right column, are built from $\deltabetadust$ and $\deltabetadeuxdust$ maps, respectively weighted by the frequency dependent ratios $ \partial f(\nu)/{\partial \beta}$ and $\partial^2 f(\nu)/\partial \beta^2$. Those maps are compared with the \citedpc\ ones (left column), where each frequency SED spatial variation map has been determined independently. The large scales are similar although obtained in different ways, and despite the fact that the initial dust maps were taken from different frequencies: 857\,GHz for \citedpc\ and 545\,GHz for this work.

\begin{figure}[ht!] 
\includegraphics[width=0.9\columnwidth]{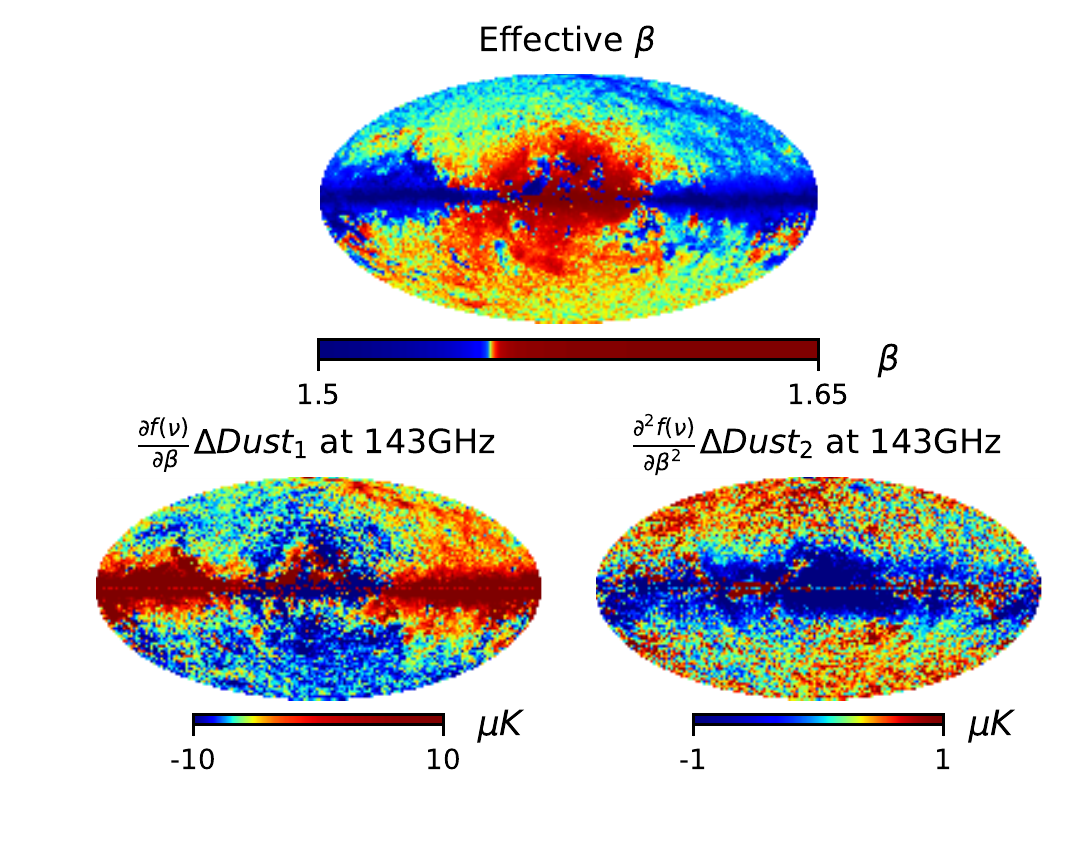} 
\centering
\caption{The top map is the effective dust parameter $\beta$ variation map. In the narrow galactic ridge, a single $\beta$ correction per line of sight is not meaningful considering the large variations of SED. The bottom maps show the two independent intensity corrections maps associated with the first and second $\beta$ order derivative.}
\label{fig:DELTA_SED} 
\end{figure}

The SED spatial variations are described for by an expansion in $\ln(\nu_b/\nu_0)$ at the second order, but associated with two independent large scales maps providing more degrees of freedom. The two bottom maps in Fig.~\ref{fig:DELTA_SED} shows these additive corrections maps for the 143\,GHz frequency band. The top map shows an estimate of an effective $\beta$ corresponding to the total correction of the two terms.

\begin{figure}[ht!] 
\includegraphics[width=0.5\textwidth]{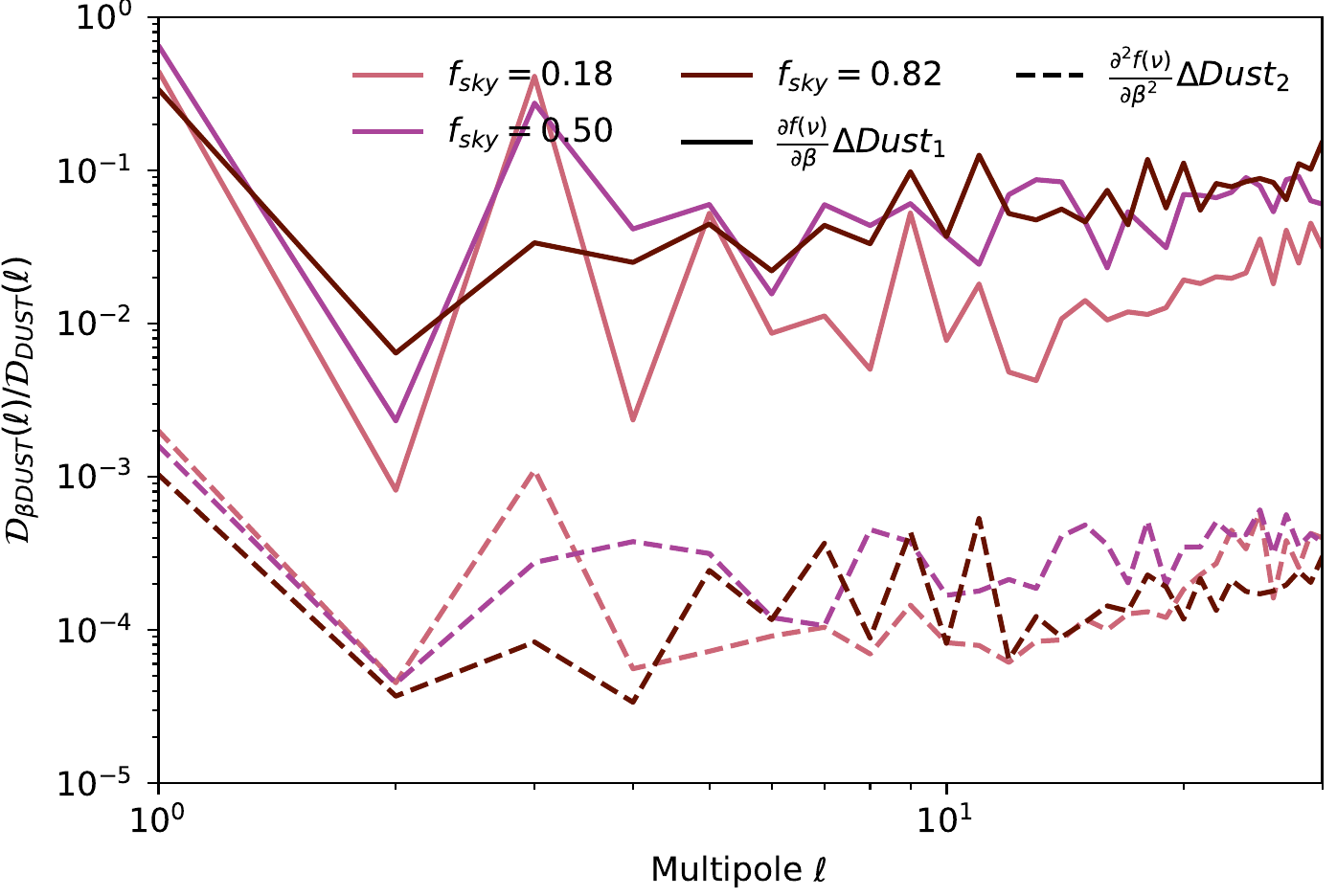} 
\caption{For different high latitude masks, the continuous lines show the power spectrum ratio of the $\deltabetadust$ term to the dust emission. The dashed lines show the power spectrum ratio of the $\deltabetadeuxdust$ term to the dust emission. These results are given for the 143\,GHz.}
\label{fig:DELTA_SED_SPEC} 
\end{figure}

To be more quantitative, Fig.~\ref{fig:DELTA_SED_SPEC} compares the ratio of the power spectra of the first and second order additive SED correction maps at 143\,GHz, with different \fsky.
There is a strong galactic center anti-center dipole term ($\ell=1$ and 3), and weaker latitude pattern ($\ell=2$ and 4), and then a white power distribution, probably reflecting the random interstellar cloud distribution. The low rise at $\ell>15$ shows the noise contribution. For \fsky=0.82, the first order SED variation corrections show a strong centre to anticentre dipole term at the 30 to 50\% level of the dust initial spectra with opposite sign to the dust map. The second order correction is two orders of magnitude smaller (one order of magnitude in the map). This agrees with the astrophysical assumption that the dust SED gradients are induced by large scale galactic physics, - gradient of gas molecular fraction in latitude and star light energy density gradient with galactic radius -, and not small scales variations of the dust properties which are seen in the flat power spectrum ($5<\ell<15$). 

To test that the dust model is valid up above 545\,GHz, we describe, in Appendix~\ref{sec:tartempion}, its extension to 857\,GHz. 

\section{CMB large scales results and solar dipole determination}
\label{sec:SolarDip_det}

The solar dipole is extracted from each single detector \CMBtot\ map from which we can subtract CMB anisotropies. The first iteration uses the \Planck\ Legacy CMB anisotropies {\tt SMICA} map, after masking the galactic ridge keeping 95\% of the sky as recommended in \cite{planck2016-l04}. In later iterations, the Bcsep \CMBtot\ map from Equation~\ref{eq:residualcmb} is used.

\subsection{New CMB large scale anisotropies map}
\label{sec:cmb-anisot-removal}

To illustrate the accuracy of the CMB cosmological anisotropies map built by the \Bcsep\ method, we use an end-to-end simulation which contains CMB, noise, systematic effects residuals, and also a CO model and a dust model as per Eq.~\ref{eq:dustmodel}. Thus, this simulation does not intend to evaluate the \Bcsep\ ability to solve for the dust model, but only for the noise and systematic effects residuals. 

\begin{figure}[ht!] 
\includegraphics[width=0.9\columnwidth]{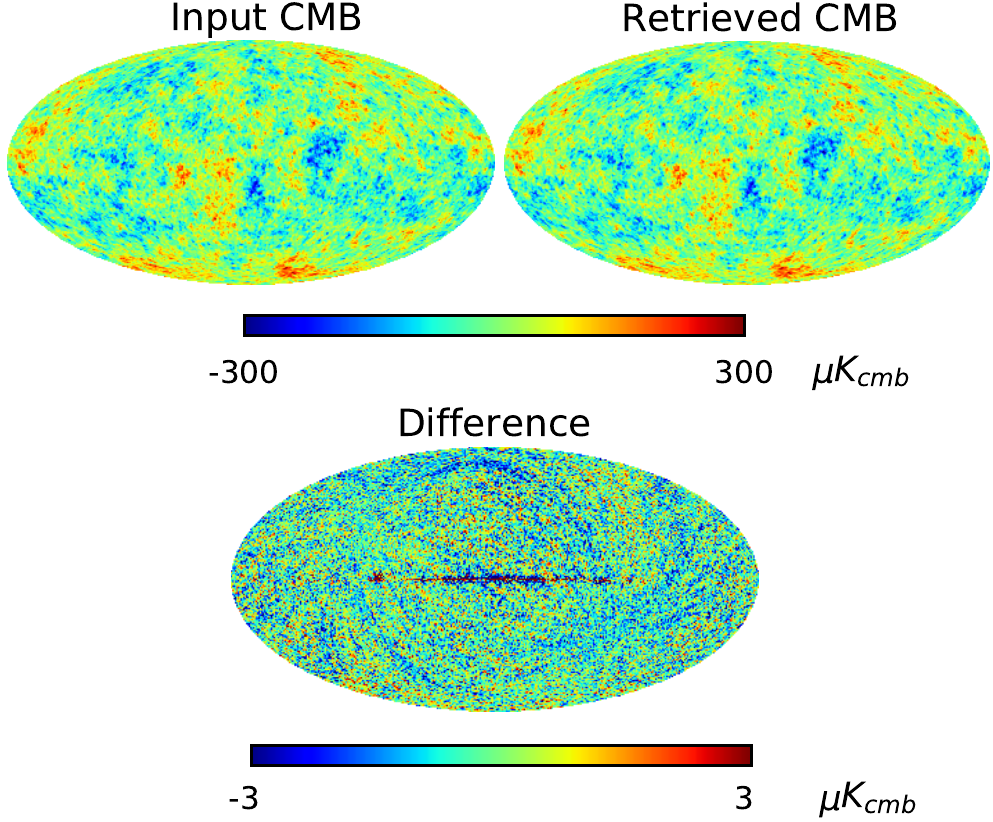} 
\centering
\caption{Retrieved primordial CMB anisotropies \Bcsep\ map compared to the input CMB (top maps). The bottom map shows the difference between input and output with a color scale 100 times smaller.}
\label{fig:CMB_SIMU} 
\end{figure}

Figure~\ref{fig:CMB_SIMU} shows the difference between the simulated input CMB primary anisotropies map (without solar dipole) and the retrieved one.
The narrow galactic ridge towards the inner galaxy is the only visible contributor to a significant dipole term, with a main vector direction toward $b=l=0\deg$. To avoid this bias, the solar dipole is extracted outside the galactic ridge mask $M_g$. The error induced on the solar dipole amplitude is 0.103\microK\ with \fsky= 0.90, and 0.108\microK\ with \fsky= 0.95. This shows that the noise and systematic effect on the CMB maps removal outside $M_g$, within this range of \fsky, is not a direct problem for the solar dipole determination. The dust foreground residuals remains the main issue.

\begin{figure*}[ht!] 
\includegraphics[width=\textwidth]{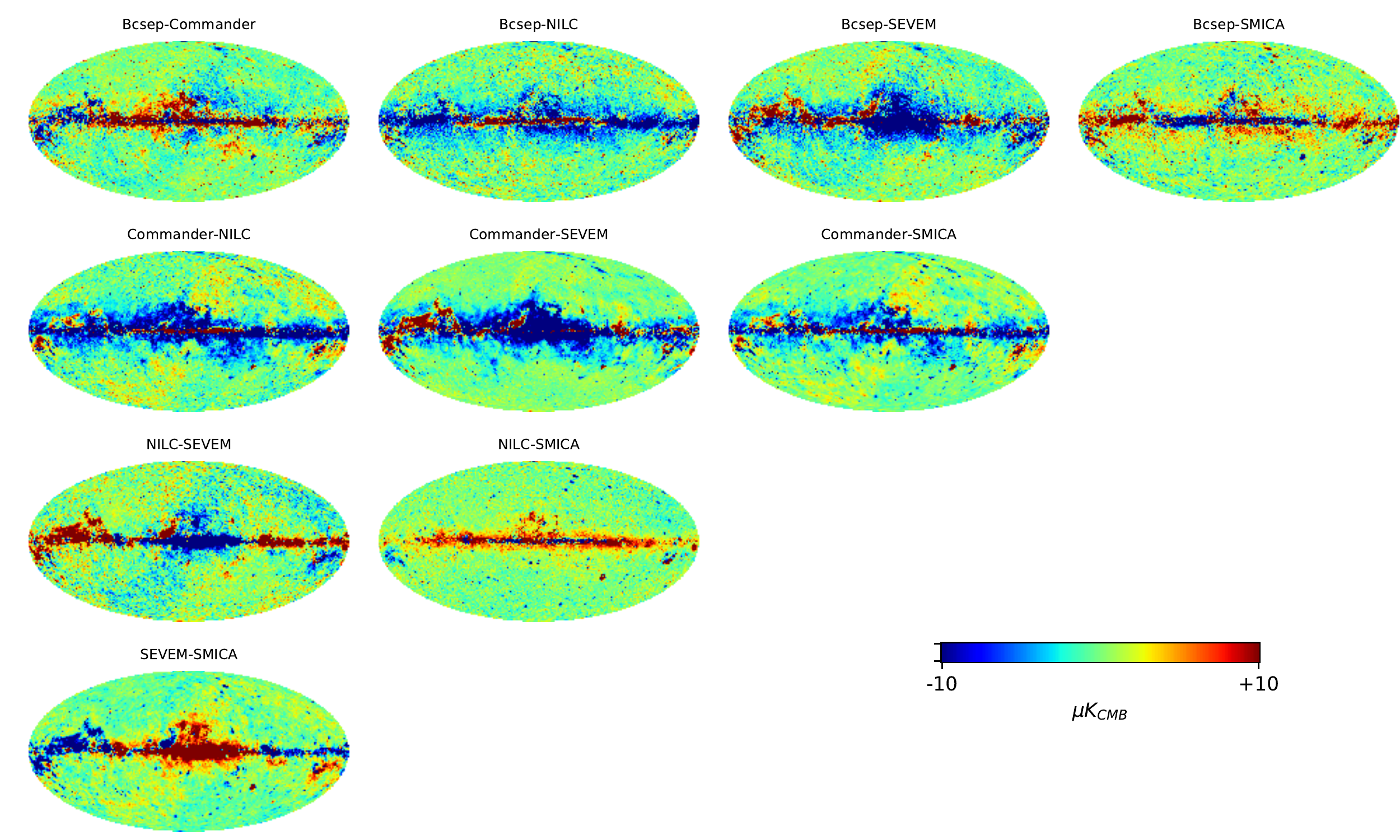} 
\caption{Difference between the CMB maps built by \Bcsep\ and the four \Planck\ Collaboration component separation methods.}
\label{fig:CMB_DIFF} 
\end{figure*}

Figure~\ref{fig:CMB_DIFF} displays the difference maps between the \CMBtot\ maps obtained by \Bcsep\ and the four \Planck\ Collaboration methods {\tt Commander}, {\tt NILC}, {\tt SEVEM}, and {\tt SMICA} \citep{planck2016-l04}.
The residuals are dominated by the galactic foregrounds at low latitudes. At high galactic latitude ($b>20\deg$), {\tt NILC}, and {\tt SMICA}, which do not depend on a physical model of the galactic components to extract the Gaussian CMB, are in very good agreement. The difference map between {\tt SMICA} and \Bcsep\ at $b>20\deg$ does not show significant residuals either. All differences containing either the {\tt Commander} or the {\tt SEVEM} maps show zodiacal dust emission residuals along the ecliptic equator.

We note that the \CMBtot\ \Bcsep\ map is only aimed at a significant improvement at $\ell<30$ and outside the bright galactic ridge $M_g$ where it has smaller large scale dust residuals than the \Planck\ legacy ones. \Bcsep\ does not intend to build a CMB map aimed at the analysis of cosmological parameters mostly based on CMB power spectra at $\ell > 30$, and is not characterized for that purpose.
 
\subsection{Optimization of the CMB anisotropies removal}
\label{sec:dipolerem}

Lacking a model to build statistically representative and controlled dust foreground maps, the foreground residuals cannot be fully evaluated by simulations. A semi-quantitative approach is nevertheless possible by comparing data and simulations. The behavior of the solar dipole parameters with frequency and sky fraction are sensitive probes to test both the CMB anisotropies removal method and the galactic foreground residuals.

Large scale foreground residuals in the part of the sky where the dipole is fitted, induce a source of systematic effects on the dipole parameter retrieval. These effects are introduced in the simulation by adding an empirical estimate of the foreground residuals. This residual is defined as the difference map from two component separation method : {\tt SMICA}-\Bcsep\, shown in Fig.~\ref{fig:CMB_DIFF}.

We now use the \CMBtot\ \Bcsep\ map built in Sect.~\ref{sec:cmb-anisot-removal}. When extracting the solar dipole from the full sky \CMBtot, the solar dipole parameters are biased by the galactic ridge residuals seen in Fig.~\ref{fig:CMB_DIFF}. This leads to the use of the galactic ridge mask $M_g$ where the \CMBtot\ is affected by large galactic residuals. This mask $M_g$ is filled with a simulated CMB constrained at the boundaries \citep{2004PhRvD..70h3511W, thommesen}. By definition, the all sky CMB cosmological anisotropies map has no dipolar term, and shall be built by removing the dipolar term from the \CMBtot\ map. The solar dipole is thus fitted in a mask $\mathcal{M}$ taken as the complement of $M_g$. On the one hand, the masked sky replaced by a constrained CMB, induces an uncertainty on the dipole parameters. This uncertainty is evaluated by the dispersion of the retreived solar dipole parameters from 1000 Monte Carlo realizations with various $M_g$. On the other hand, the bias, induced by the dust residuals in the central galactic ridge, decreases when increasing $M_g$. The combination of these two effects presents a minimum, which defines the optimal. The large scale dust residuals in the  \CMBtot\ map also affect slightly the dipole measurement (Sect.~\ref{sec:cmb-anisot-removal}). Any detectable dust residual effect will then appear as a frequency and/or an \fsky\ dependence of the dipole parameters when changing $M_g$.

\begin{figure*}[ht!] 
\includegraphics[width=0.9\textwidth]{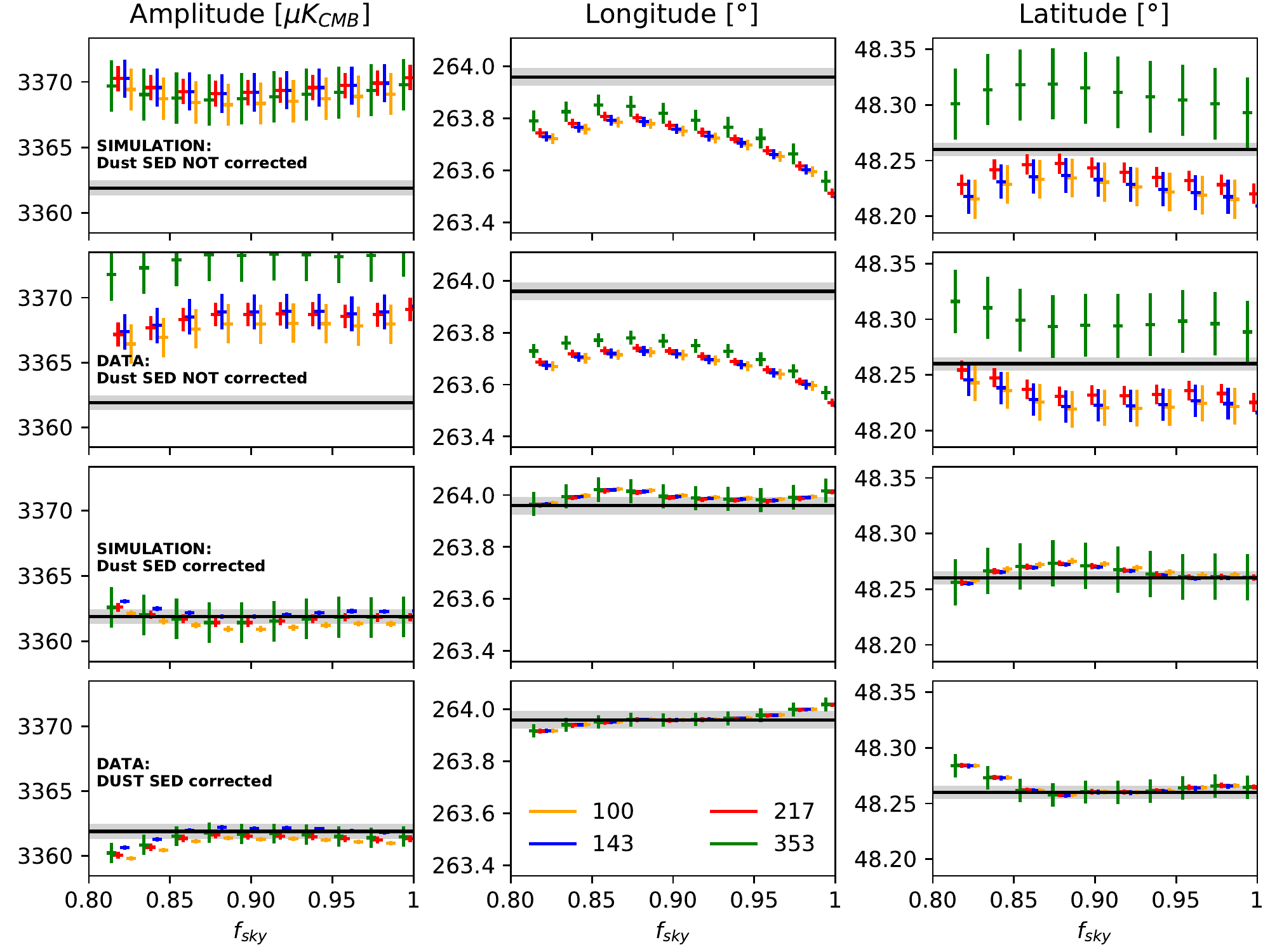} 
\centering
\caption{Reconstructed solar dipole parameters along sky fraction used to remove the galactic ridge from the retrieved CMB. Frequencies are color coded from 100 to 353\,GHz. Black lines reflect the input dipole parameters (taken from the final result of this work). The grey band gives the uncertainties on these parameters.}
\label{fig:COMSEP_BIAS_DIPOLE} 
\end{figure*}

We want to compare these two effects: the large scale SED spatial variations of the dust emission, and the \CMBtot\ filling of $M_g$. These behaviors, observed with the data, are simulated. Figure~\ref{fig:COMSEP_BIAS_DIPOLE} shows the computed solar dipole parameters for the 100-353\,GHz frequencies, when varying the galactic ridge mask $M_g$ in which the \CMBtot\ \Bcsep\ map is filled with constrained CMB realizations. The dipole parameters are plotted as a function of the sky fraction left outside $M_g$ used to fit the dipole.

At first, the systematic effects affecting the retrieved dipole parameters are tested both on simulations and data to ensure that the simulations are a good representation of the data. The two first rows if Fig.~\ref{fig:COMSEP_BIAS_DIPOLE} use the processing algorithm without SED variation correction. The input solar dipole parameters for the simulated data are taken from the final determination from this work, and given by the black lines with grey bands. In these two first rows, we see the remaining effects of the dust SED variations effects (included in the simulations and not corrected for) well simulated when compared to the results obtained on the data. The magnitude and sign of the residuals are qualitatively demonstrating the quality of the simulation even though the dust behavior is not quantitatively exact as expected, lacking a reliable physical model of the dust SED variations. In the third and fourth rows of Fig.~\ref{fig:COMSEP_BIAS_DIPOLE}, the algorithm now includes the dust SED spatial variation correction. Both show a spectacular reduction of the previous trends. This demonstrates that the simulation takes into account the SED variations with the right order of magnitude, and that also the algorithm captures them well. 

\begin{figure}[ht!] 
\includegraphics[width=0.90\columnwidth]{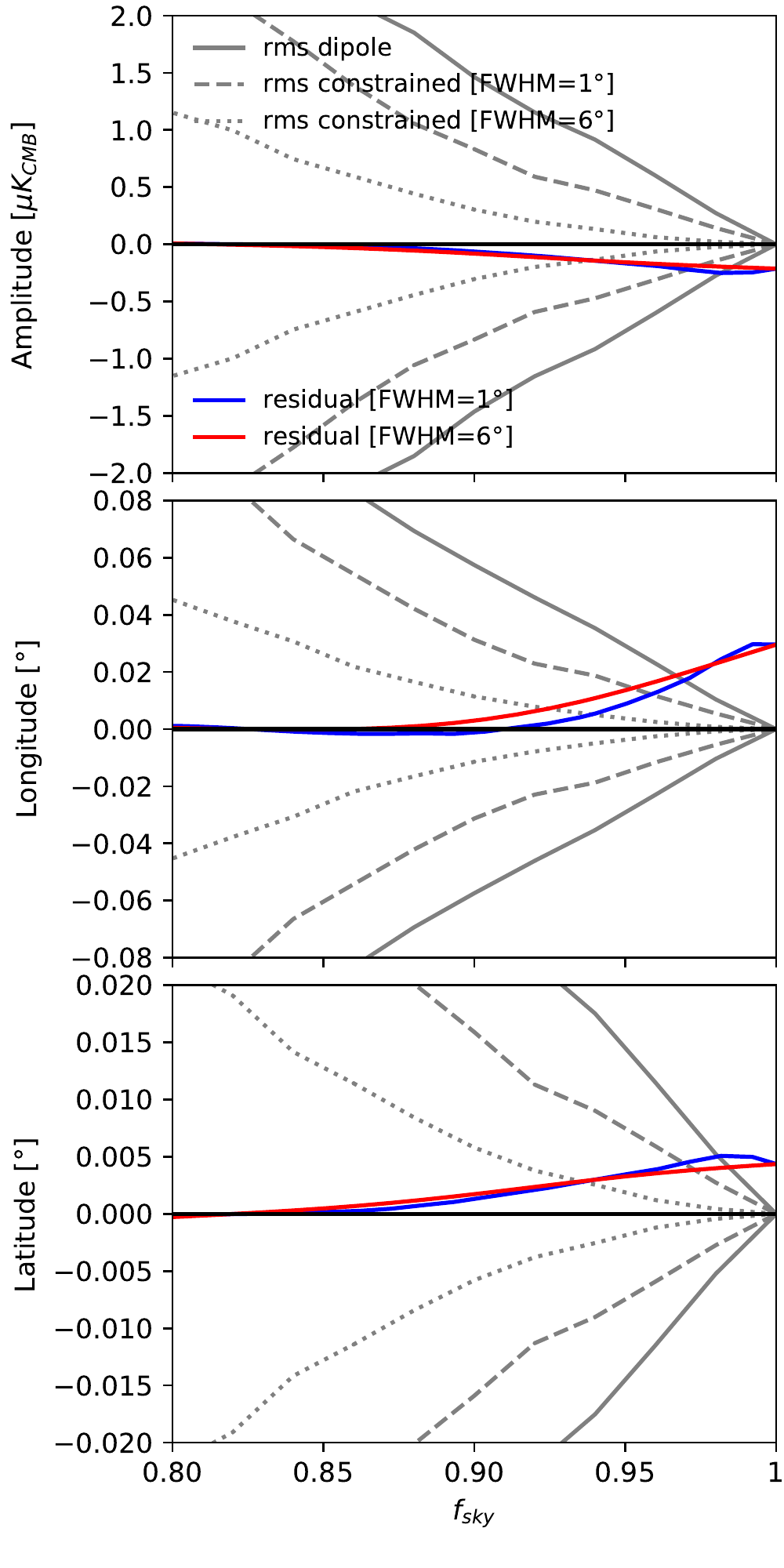} 
\centering
\caption{The retrieved solar dipole parameters as a function of the sky fraction outside the galactic ridge mask $M_g$ out of which the solar dipole is extracted. $M_g$ is filled with a contrained realization of CMB anisotropies. The blue and red lines show the bias for two smoothing (1\deg and 6\deg). The gray wedges show the dispersion of the solar dipole parameters.}
\label{fig:themask}
\end{figure}

The dipole term has to be removed from the \CMBtot\ map, well known only outside a galactic ridge mask $M_g$ where there are significant galactic dust residuals. This masking induces an error due to leakages between the very large scale anisotropies and the dipole term. Figure~\ref{fig:themask} shows the results on the recovered parameters of the solar dipole as a function of the sky fraction left outside of the varying $M_g$ where the dipole is fitted. 
The dispersion of the solar dipole parameters is evaluated by simulation with 1000 realizations of the CMB filling of $M_g$. The plain gray wedges show the dipole term dispersion when nothing is done to compensate this leakage. The dashed (resp. doted) gray wedges show the dispersion of the dipole term after filling $M_g$ with a smoothed 1\deg (respectively 6\deg) CMB realization of the large scales only (up to $\ell=5$) constrained by the cosmological parameters HFI best-fit \citep{planck2016-l05}, and continuity constraints at the boundaries. Once the dipole term is removed from the \CMBtot\ map, as per the adopted definition, it is used to remove the CMB anisotropies from the full sky map. Then, the solar dipole is fitted in $\mathcal{M}$. This removal and filling procedure introduces a non-recoverable uncertainty, increasing with $M_g$ estimated using results shown in Fig.~\ref{fig:themask}.

The simulations with the qualitatively representative dust component residuals defined above, show a bias as expected in the amplitudes and in direction (longitude bias as expected larger than the latitude one) of the solar dipole parameters. The biases are maximal when the galactic ridge mask $M_g$ is null, and decreases with decreasing \fsky\ following the predicted behavior. For no masking of the galactic ridge, the direction is shifted by about 2\arcmin\ of longitude. When \fsky\ decreases to 0.95, the longitude bias is reduced by a factor of 2, and crosses the longitude dispersion for 1\deg smoothing. For this sky fraction, the amplitude parameter shows a bias of 0.25\microK. The dispersion for $6\deg$ smoothing crosses the bias for \fsky=0.92. We thus conclude that, for the CMB anisotropies removal map, a $M_g$ mask leaving \fsky=0.95 available for the fit of the dipole is optimal for the CMB anisotropies removal with a safer 1\deg smoothing. 

\subsection{Dipole fit stability}
\label{sec:dipole_stability}

\begin{figure}[ht!] 
\includegraphics[width=0.85\columnwidth]{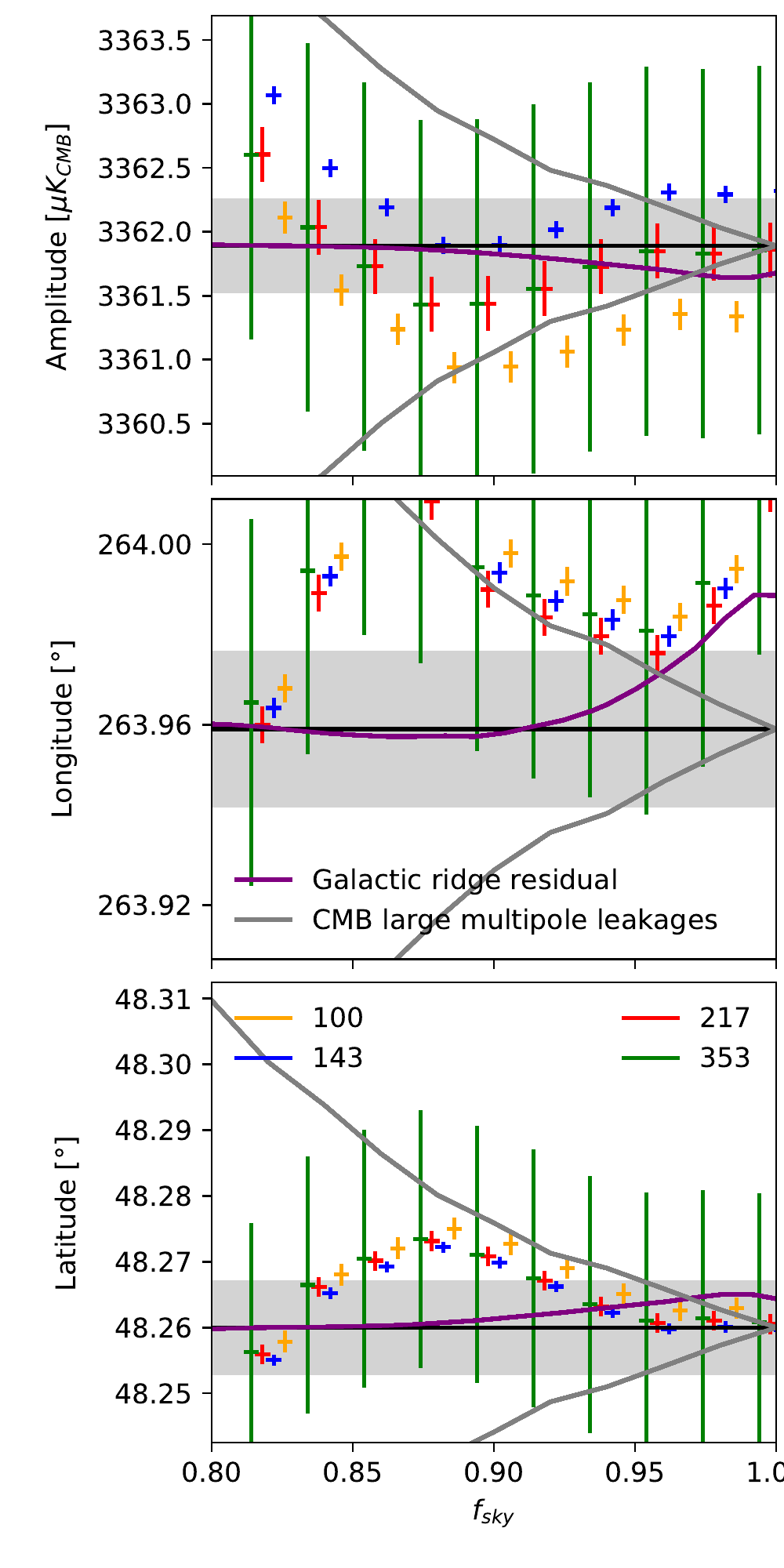} 
\centering
\caption{Zoom-in of the Fig.~\ref{fig:COMSEP_BIAS_DIPOLE} for simulated data. The purple lines show the galactic ridge emission residual after the \Bcsep\ component separation. The gray lines show the dispersion of the effect of $M_g$ filling.}
\label{fig:DIP_DATA} 
\end{figure}

Figure~\ref{fig:DIP_DATA} is a zoom-in of Fig.\ref{fig:COMSEP_BIAS_DIPOLE} third row.
The dipole parameters for all frequencies show the behavior described in Sect.~\ref{sec:dipolerem} with a bias increasing for increasing from \fsky=0.8 to 1.0 (purple line), and a dispersion, as the gray wedge ($1\deg$ smoothing), decreasing rapidly with increasing \fsky. The four frequencies show very similar patterns. The longitude, the most sensitive parameter, shows a minimal error for the optimal $M_g$ size to \fsky=0.95 as expected. The behavior of the three parameters shows a correlated, but non monotonic, behavior as expected when dominated by the dispersion induced by the large scales foregrounds for $\fsky < 0.95$, within the range predicted by the gray wedge. It is at present impossible to draw high quality statistically representative galactic foreground maps, and thus fully evaluate the foreground residuals. The uncertainties are given by the dispersion between detectors in each frequency band, and increase with frequency. Calibration errors resulting from noise and systematic effects, both evaluated by simulations, are visible as a variation in amplitude between frequencies. 

\subsection{Solar dipole results}
\label{sec:dipoleresults}

We want now to test the effect of the dust residuals on the stability of the solar dipole parameters, when increasing frequencies, and varying $\mathcal{M}$ keeping the optimal $M_g$ constant (\fsky=0.95). 

\begin{figure*}[h] 
\includegraphics[width=0.90\textwidth]{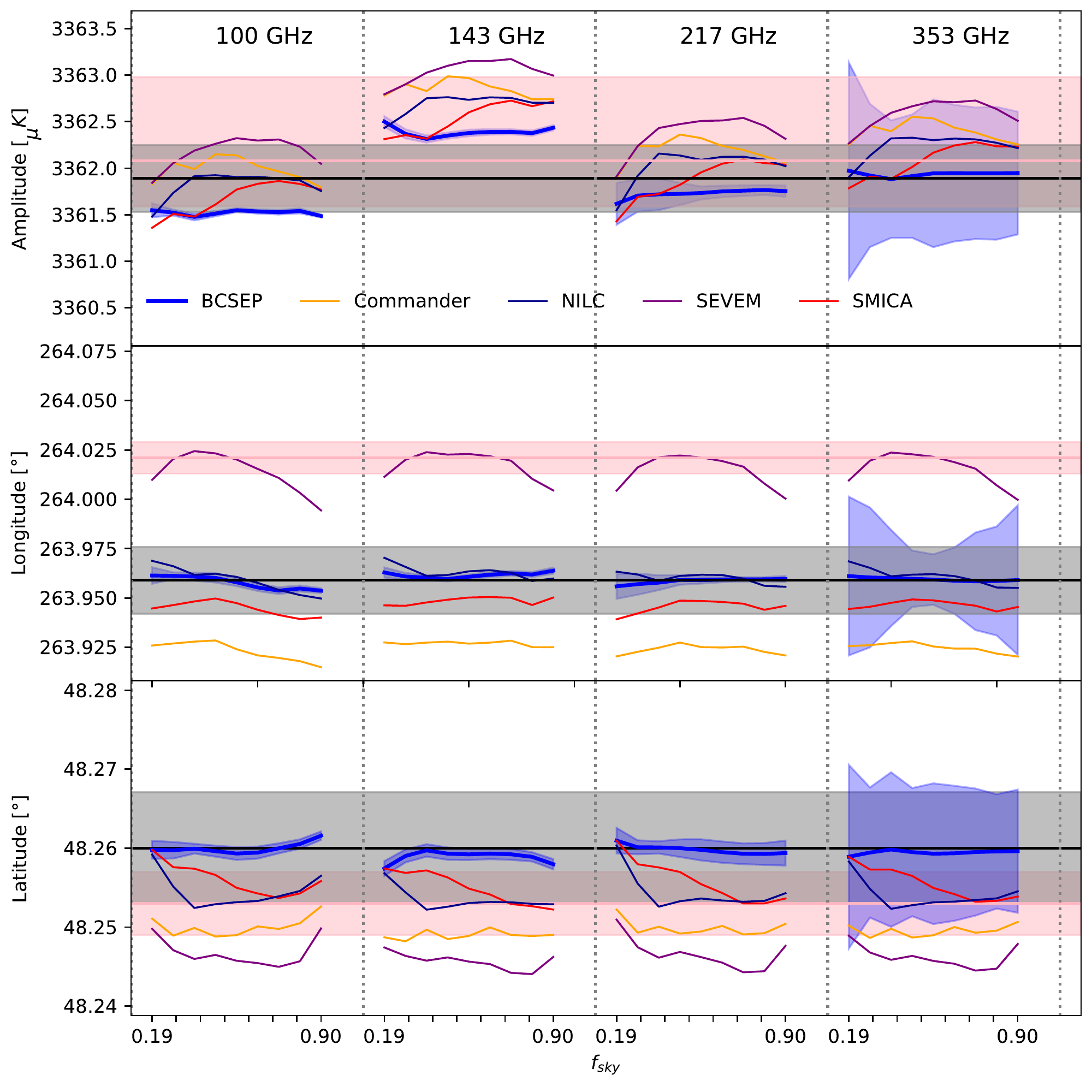} 
\centering
\caption{Variation of the solar dipole parameters (in rows) for different frequency bands (in columns) as a function of the galactic mask in which the dipole is fitted. The black lines and grey band uncertainties show the final results of this work, using the \Bcsep\ CMB anisotropies map. For reference, the pink lines and bands show the \citedpc\ measurements. The four \Planck\ Legacy CMB anisotropies maps are also tested, and are shown color coded.}
\label{fig:fskylargevar} 
\end{figure*}

Figure~\ref{fig:fskylargevar} shows the stability of the solar dipole parameters when the fitting mask $\mathcal{M}$ spanning from \fsky=0.19 to 0.9 (see Sect.~\ref{sec:dipolerem}).
The solar dipole parameters stability is impressive, and do not show any obvious trend with the dipole fitting mask $\mathcal{M}$ nor with frequency. The most noticeable discrepancies are in amplitude and nearly all within $1\sigma$. The final results of this paper (average over the three lower frequencies) is shown as black lines and grey band uncertainties. We thus conclude that there is no dust foreground residuals detectable above the noise.

Table~\ref{tab:dipoleresults} shows the final results of the solar dipole parameters, computed as the weighted averaged per frequency over $\mathcal{M}$ with \fsky=0.3 to 0.8.
\begin{table}[h] 
\newdimen\tblskip \tblskip=5pt
\caption{Solar dipole parameters averaged per frequency and per fitting mask $\mathcal{M}$. AVG is the average over CMB dominated frequencies 100, 143, and 217\,GHz. The 353\,GHz is also given; it has larger uncertainties because of the much larger dust emission.}
\label{tab:dipoleresults}
\vskip -6mm
\footnotesize
\setbox\tablebox=\vbox{
\newdimen\digitwidth
\setbox0=\hbox{\rm 0}
\digitwidth=\wd0
\catcode`*=\active
\def*{\kern\digitwidth}
\newdimen\signwidth
\setbox0=\hbox{+}
\signwidth=\wd0
\catcode`!=\active
\def!{\kern\signwidth}
\halign{\hbox to 1.cm{#\leaderfil}\tabskip 1em& \hfil$#$\hfil\tabskip 1em& \hfil$#$\hfil\tabskip 1em& \hfil$#$\hfil\tabskip 0em\cr
\noalign{\doubleline}
\omit\hfil Frequency\hfil&A&l&b\cr
\omit\hfil [GHz]\hfil&[\muK]&$[deg]$&$[deg]$\cr
\noalign{\vskip 3pt\hrule\vskip 5pt}
100& 3361.52 \pm 0.04 & 263.958 \pm 0.003 & 48.260 \pm 0.001 \cr
\noalign{\vskip 4pt}
143& 3362.36 \pm 0.04 & 263.961 \pm 0.002 & 48.259 \pm 0.001 \cr
\noalign{\vskip 4pt}
217& 3361.74 \pm 0.11 & 263.959 \pm 0.002 & 48.260 \pm 0.001 \cr
\noalign{\vskip 4pt}
\textit{353}&\textit{ 3361.93 $\pm$ 0.71} & \textit{263.959 $\pm$ 0.019} & \textit{48.259 $\pm$ 0.009} \cr
\noalign{\vskip 4pt}
AVG& 3361.90 \pm 0.36 & 263.959 \pm 0.003 & 48.260 \pm 0.001 \cr\noalign{\vskip 4pt}
\noalign{\vskip 3pt\hrule\vskip 5pt}}}
\endPlancktablewide
\end{table}
For amplitudes, the dispersion between detectors within a frequency band is a more relevant measure of the whole calibration process, including systematic effects.\citedpc\ uses 100 full end-to-end simulations of the overall absolute calibration process based on the recovery of the solar dipole input. These simulations retrieve the rms uncertainties of the calibration measured by the solar dipole amplitude rms: $1.5 \times 10^{-4}$ at frequency from 100 to 217\,GHz increasing to $4\times 10^{-4}$ at 353\,GHz. Our results in amplitude are coherent with the \citedpc\ calibration mismatch estimates. Thus, the weighted average (AVG) of the 100, 143, and 217\,GHz results provides the best dipole parameters estimate. The AVG amplitude uncertainty (0.36\microK), computed as the dispersion between the three frequencies, is significantly larger than the intra-frequency uncertainties, but reflects the calibration error due the residual systematic effects. 

For direction, the uncertainties are the statistical errors. The \citedpc\ results are obtained by alignment of the 143-353\,GHz frequencies solar dipole directions on the 100\,GHz solar dipole direction, assuming it was the best reference, because of its best combination of high CMB sensitivity and low foregrounds. In the present work, this assumption is replaced by the constraint of minimum dispersion of the dipole directions for all 100 to 353\,GHz detectors, a more refined dust model of the SED spatial variation, and coherent large scales CMB anisotropies. We also introduce a better masking-filling procedure for the removal of the galactic ridge residuals in the CMB cosmological anisotropies. Figure~\ref{fig:fskylargevar} compares the present results with the \citedpc\ results, shown as the pink lines and bands. The results for the amplitude and latitude are well compatible within their error bars. Nevertheless, the longitude is not, although very stable when varying \fsky, and consistent with {\tt SMICA}, and {\tt NILC}. 

An empirical test of the systematic effects related to the component separation is to use different CMB maps. This is done also in Fig.~\ref{fig:fskylargevar}, and the four component separation method used for the CMB anisotropies are shown with thin lines. Although using CMB from {\tt Commander}, {\tt NILC}, {\tt SEVEM}, and {\tt SMICA} is not fully consistent with the new and better foreground dust model developed in this work, the comparison remains interesting. When other component separations are used to get the CMB anisotropies, the stability with varying \fsky($\mathcal{M}$) parameter is not as good as the one achieved by the \Bcsep\ one. It strengthens the statement that the CMB removal, when done with an optimal galactic ridge mask, has little effect on the solar dipole parameters fitted in a mask $\mathcal{M}$ covering a very broad range of sky fraction, thus showing no sign of large scales dust residuals. 
 
\cite{thommesen} and \cite{npipe} increasing the galactic mask up to 80\% of the sky, is far from optimal. The dispersion of the amplitude in these papers grows from $\pm2.5$\microK\ for $M_g$ in the range \fsky\ 0.95 to 0.75 and reach $\pm10$\microK\ for \fsky=0.30. These large uncertainties are induced when setting the dipole term to zero in a larger and larger galactic mask. Conversely, the method described in the present work, optimizes this mask and reduce by large factors the uncertainties making full use of the CMB anisotropies accurately measured up to 95\% of the sky. We also note that the $4 \sigma$ calibration discrepancy between 100 and 143\,GHz mentioned in Table~10 if \cite{npipe} is of the same order as the absolute calibration reflected by the shift of the amplitude of the solar dipole by 4\microK\ in their paper showing a photometric calibration problem.

\begin{figure}[h] 
\includegraphics[width=0.9\columnwidth]{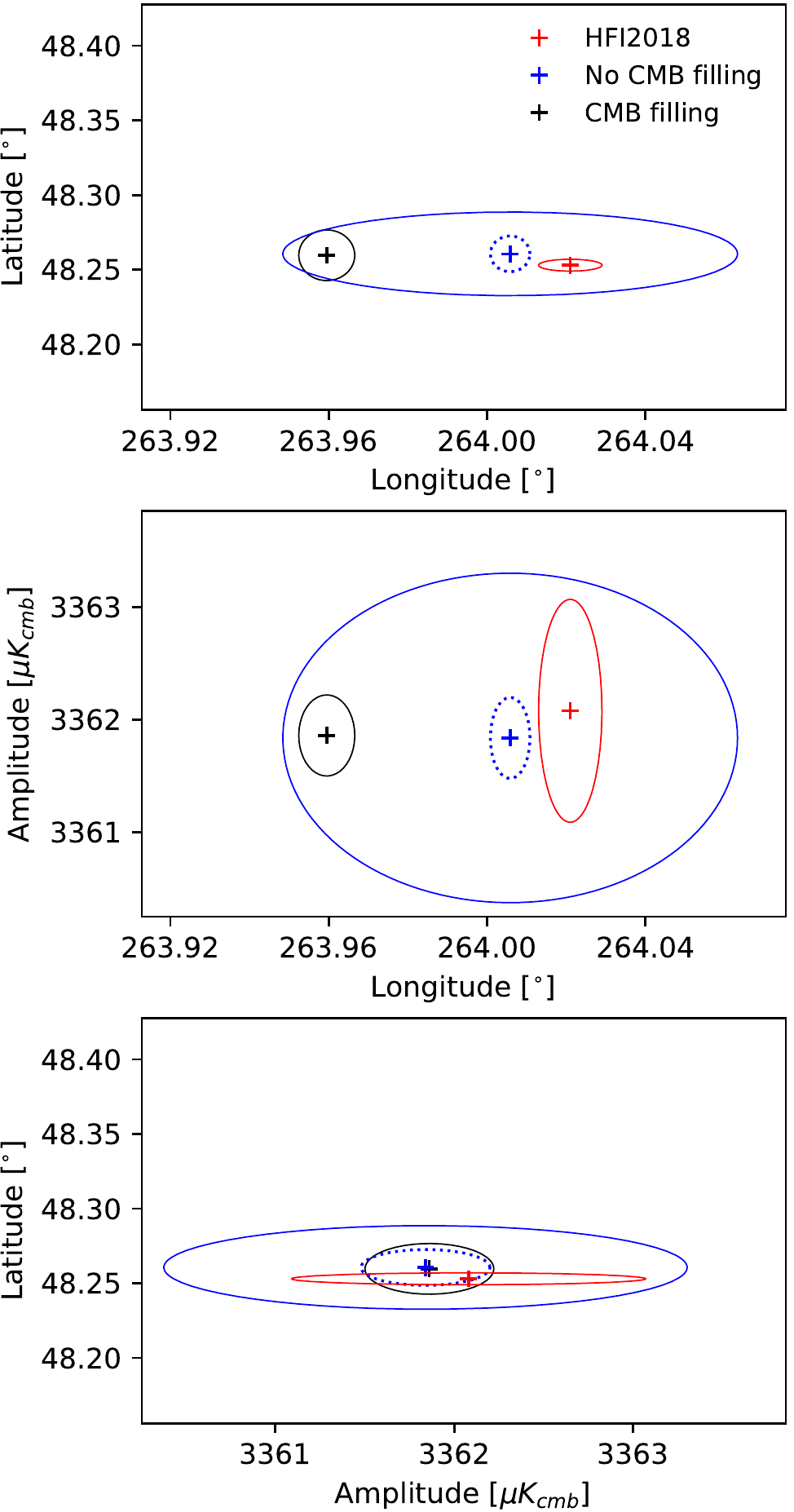} 
\centering
\caption{Three combinations of the solar dipole parameters as a cross together with uncertainty ellipses. Red color shows the \citedpc\ results. The blue color is the present work but without filling the galactic ridge mask $M_g$, dotted when ignoring the systematic effect on the CMB anisotropies dipolar term, full line when taking into account the systematic effect. The black color is for the results of the present work with galactic ridge $M_g$ filled with a constrained CMB.}
\label{fig:DIP_DATA_RES2} 
\end{figure}

Figure~\ref{fig:DIP_DATA_RES2} shows three combinations of the solar dipole parameters, results of this work, and of the \citedpc\ results (in red).
The longitude is the only discrepant parameter, being shifted by $- 3.6\arcmin$ ($2.5\arcmin$ on the sky) well outside the error ellipses. There is a key difference between the extraction method used in this work and the one used in \citedpc. The latter fit the dipole parameters outside of a galactic ridge mask for which the zero dipolar term of the CMB anisotropies has been removed without filling the mask with constrained CMB realizations non contaminated by dust residuals. This induces a bias on the dipolar component in the part of the sky outside the galactic ridge mask $M_g$. This bias has been simulated on the present final maps, by removing the step of the gap filling, and fitting the dipole parameters outside the optimal mask $M_g$. The dipole direction is displayed as the blue cross and dotted error ellipse in the top panel of Fig.~\ref{fig:DIP_DATA_RES2}. This point falls near the \citedpc\ red point, as expected as both of them have not the optimal gap filling. The bias on the sky is $1.8\arcmin$ when the shift observed is $2.5\arcmin$. This simulation shows the bias induced when $M_g$ is not filled with a constrained CMB, and we added this bias to the uncertainties of the legacy shown by the full blue line ellipse.

Averaging the three lowest Planck-HFI CMB frequencies (Table~\ref{tab:dipoleresults}) with uncertainties including the systematic effects leads thus to the final solar dipole parameters:
\begin{eqnarray}
A&=& \left[3361.90\pm0.04\phantom{0}\,\text{(stat.)}\pm0.36\phantom{0}\,\text{(syst.)}\right]\,\mu{\rm K}; \nonumber\\
l&=&\text{263\pdeg959} \pm \text{0\pdeg003}\,\text{(stat.)} \pm \text{0\pdeg017}\,\text{(syst.)}; \nonumber\\
b&=&\text{\phantom{0}48\pdeg260} \pm \text{0\pdeg001}\,\text{(stat.)}\pm \text{0\pdeg007}\,\text{(syst.)}.
\end{eqnarray}
This result is consistent with \citedpc\ for amplitude and latitude. The longitude of the dipole axis is shifted by $-2.5\arcmin$ on the sky. This is fully explained by the difference of procedure, having introduced the filling of the galactic ridge mask with constrained CMB anisotropies realizations before setting its dipolar term to zero. 

\section{Conclusion}

This work provides the best measurement of the Solar dipole parameters as reminded in Table~\ref{tab:dipolehistory}, showing an unprecedented stability. This measurement, together with the large scale CMB cosmological anisotropies map can be used for intercalibration of future CMB experiments, and testing the removal of foregrounds as done in this work.
\begin{table*}[h] 
\newdimen\tblskip \tblskip=5pt
\caption{Measurements of the solar dipole parameters for different Collaborations and data sets.}
\label{tab:dipolehistory}
\vskip -6mm
\footnotesize
\setbox\tablebox=\vbox{
\newdimen\digitwidth
\setbox0=\hbox{\rm 0}
\digitwidth=\wd0
\catcode`*=\active
\def*{\kern\digitwidth}
\newdimen\signwidth
\setbox0=\hbox{+}
\signwidth=\wd0
\catcode`!=\active
\def!{\kern\signwidth}
\halign{\hbox to 4.6cm{#\leaderfil}\tabskip 0.5em&$#$\hfil\tabskip 0.5em&$#$\hfil\tabskip 0em&$#$\hfil\tabskip 0.5em&$#$\hfil\tabskip 0em&$#$\hfil\tabskip 0.5em&$#$\hfil\tabskip 0em&$#$\hfil\tabskip 0em\cr
\noalign{\doubleline}
\omit\hfil Reference\hfil&\hfil\text{Collaboration data set}\hfil&\multispan2 \hfil Amplitude\hfil &\multispan2 \hfil Longitude \hfil &\multispan2 \hfil Latitude \hfil\cr
\omit\hfil & &\multispan2 \hfil[\muK]\hfil &\multispan2 \hfil[deg]\hfil&\multispan2 \hfil[deg]\hfil \cr
\noalign{\vskip 3pt\hrule\vskip 5pt}
\cite{1996ApJ...473..576F}&\text{COBE FIRAS}&3372**&\pm7*** &264.14*&\pm0.30 & 48.26*&\pm0.30\cr
\cite{2009ApJS..180..225H}&\text{WMAP}&3355**&\pm8& 263.99*&\pm0.14& 48.26*&\pm0.03\cr
\cite{planck2014-a01}&\text{Planck HFI+LFI}&3364.5*&\pm2.0 & 264.00*&\pm0.03 & 48.24*&\pm0.02\cr
\cite{planck2016-l03}&\text{Planck HFI}&3362.08&\pm0.09\,\text{(stat.)}& 264.021 &\pm0.003\,\text{(stat.)} & 48.253 &\pm0.001\,\text{(stat.)}\cr
\omit & & &\pm0.45\,\text{(syst.)} & &\pm0.008\,\text{(syst.)}& &\pm0.004\,\text{(syst.)}\cr
\omit & & &\pm0.45\,\text{(cal.)} & & & &\cr
This work & \text{Bware HFI} &3361.90&\pm0.04\,\text{(stat.)}&263.959&\pm0.002\,\text{(stat.)}&48.260&\pm0.001\,\text{(stat.)}\cr
\omit & & &\pm0.36\,\text{(syst.)} & &\pm0.017\,\text{(syst.)}& &\pm0.007\,\text{(syst.)}\cr
\noalign{\vskip 3pt\hrule\vskip 5pt}}}
\endPlancktablewide
\end{table*}

The dust emission is the dominant large scale anisotropies component above 100\,GHz affecting strongly the CMB at the largest angular scales. Such non-Gaussian foreground critically depends on the zero level of the frequency maps which can only be constrained for differential experiments by the coherence of the dipoles. We have developed here a method which does improve these zero levels which have much better accuracy than the extrapolation to zero column density of interstellar gas. Moreover, the component separation between the CMB and the dominant dust foreground is performed iteratively and allows to build intermediate and large scales maps of the dust SED variations. 

The data products associated to this paper are thus:
\begin{itemize}
\item improved zero levels of the HFI single detector maps,
\item refined CMB solar dipole vector parameters,
\item a new CMB anisotropies map at large scales,
\item 100 to 857\,GHz dust intensity maps of our new dust model including the SED spatial variation properties
\end{itemize}

\begin{figure}[h] 
\includegraphics[width=0.9\columnwidth]{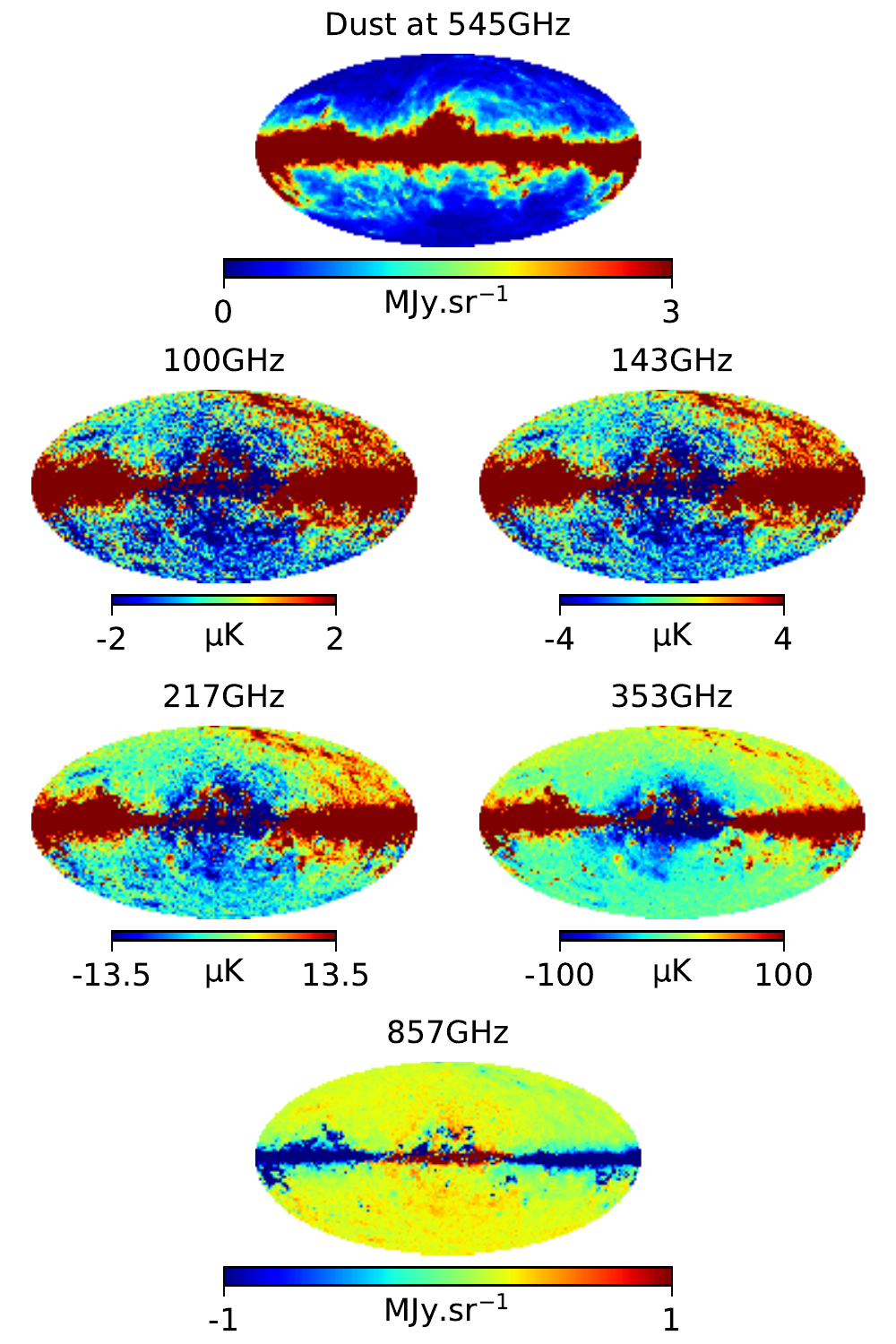}
\centering
\caption{Maps of the dust model. The top map shows the $Dust$ template extracted from the 545GHz map. The other maps show the spatial variation of the dust SED.}
\label{fig:PRODUCT} 
\end{figure}

Figure~\ref{fig:PRODUCT} illustrates this dust model, which can be used to better understand the polarized dust emission, that should be affected by similar spatial variation.

The \Planck\ HFI data will stay, for at least a decade, the best all sky data at the high frequencies. At frequencies higher than 200\,GHz, the sky signal is strongly affected by the atmosphere. The present study products are thus unique and useful to combine with balloon borne and ground based CMB data waiting for the next CMB space mission. 

\begin{acknowledgements}
This work is part of the Bware project supported by CNES. The authors acknowledge the heritage of the Planck-HFI consortium regarding data, software, knowledge. The program was granted access to the HPC resources of CINES (http://www.cines.fr) under the allocation 2020-A0080411364 made by GENCI (http://www.genci.fr). The authors thank Sylvain Mottet for producing the simulations, and Manuel Lopez-Radcenco for improving the wording.
\end{acknowledgements}

\bibliographystyle{aat}
\bibliography{solar_dipole}


\appendix 
\section{Consistency of the dust model extended to 857\,GHz}
\label{sec:tartempion}

Figure~\ref{fig:dipresidual} shows similarity between the SED correction maps from the \citedpc (based on the 857\,GHz map) and the present dust model (based on the 545\,GHz). This suggests that the dust model can be extrapolated to 857\,GHz.

The 857\,GHz map is the sum of the dust emission, an unknown offset correction, noise and systematics. The CMB anisotropies and low frequency foregrounds are fully negligible here. Thus, the 857\,GHz map is used as a dust template which is propagated at 545\,GHz using Eq.~\ref{eq:finaldustmodel}:
\begin{eqnarray} 
\mathcal{D'} (\nu_{545}) & = & \frac{f(\nu_{545})}{f(\nu_{857})} \times \\ \nonumber
&& \left[ Dust_{\nu_{857}} - \ln\left(\frac{\nu_{857}}{\nu_{545}}\right) \deltabetadust - \ln^2 \left(\frac{\nu_{857}}{\nu_{545}}\right)\deltabetadeuxdust\right]
\end{eqnarray}
A 545\,GHz dust model $\mathcal{D'} (\nu_{545})$ is built from the 857\,GHz map by propagating the SED variation dust model. This model is removed from the 545\,GHz single detector map, only leaving a CMB solar dipole map plus noise and systematics residuals. To constrain the 857\,GHz map offset $O_{857}$, we compute, for each single detector 545\,GHz map, the solar dipole parameters following the method described in Sect.~\ref{sec:SolarDip_det}. The 857\,GHz offset is adjusted by minimizing the difference between the 545\,GHz solar dipole directions and the 100-217\,GHz average (Table~\ref{tab:dipoleresults}), and found to be $O_{857}$=0.124\MJysr.

\begin{figure}[ht!] 
\includegraphics[width=0.9\columnwidth]{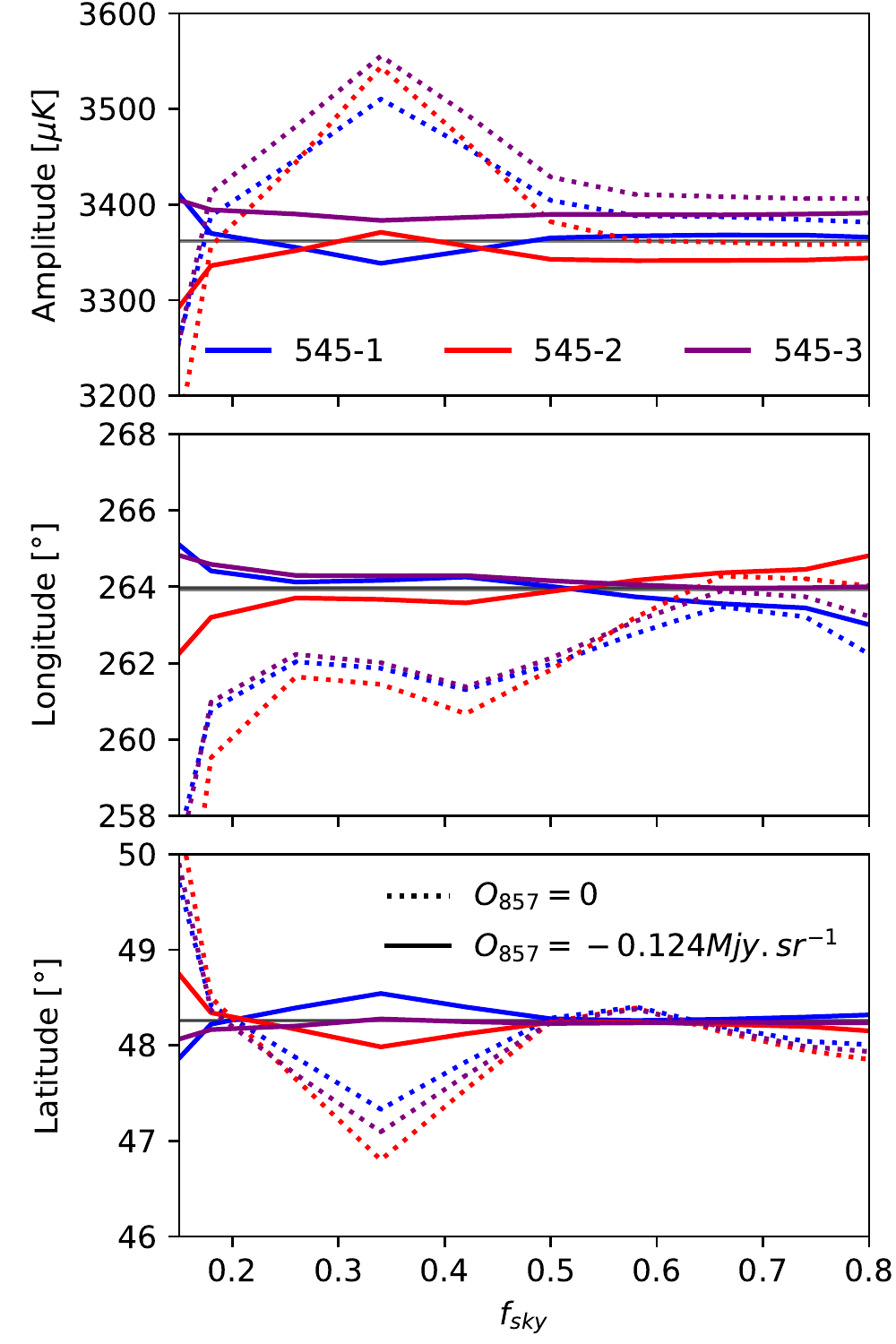} 
\centering
\caption{Solar dipole parameters extracted from single detector maps at 545\,GHz to which a dust map from 857\,GHz has been removed. The parameters are shown for the three 545\,GHz bolometers, and an offset correction null or optimized to minimize the difference with solar dipole direction.}
\label{fig:545DIPOLE} 
\end{figure}

Figure~\ref{fig:545DIPOLE} shows the CMB dipole parameters extracted from the three single detector 545\,GHz maps for both a null 857\,GHz offset and for the solved one.
The convergence and stability (amplitude dispersion about 50\microK) at such high frequencies demonstrate the quality of the dust model and the fact that the deviations do not come from the dust removal (like in the \citedpc) but more from the photometric calibration errors associated with instrument systematic effects.
 
\begin{figure}[ht!] 
\includegraphics[width=\columnwidth]{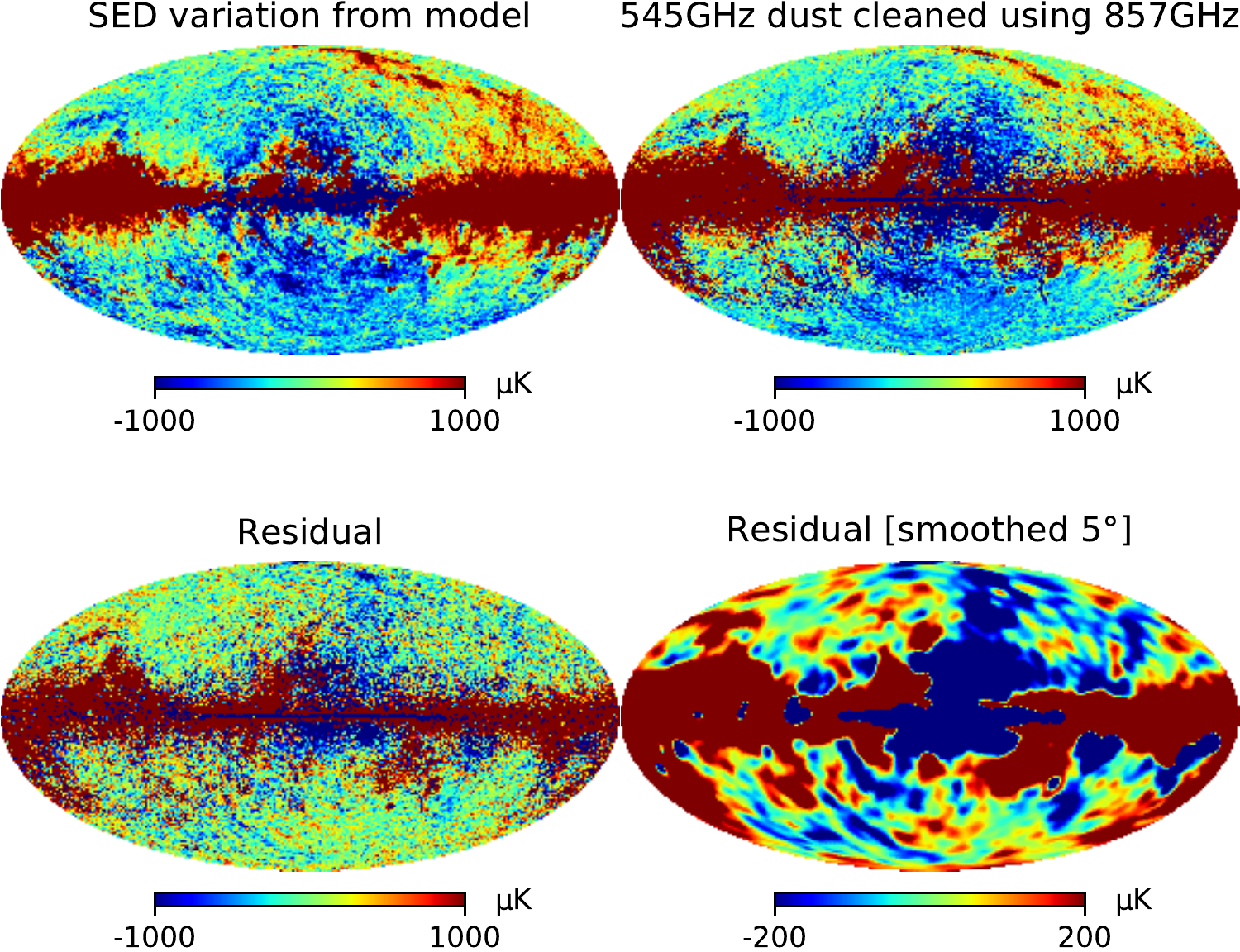} 
\caption{Top left map is the computed emission associated to the SED dust spatial variation applied to 545\,GHz. Top right is the 545\,GHz map correction after subtracting linearly the 857\,GHz map. The bottom maps are the differences between the two top maps, smoothed at 1 and 5\deg. Maps are in \microKCMB\ to allow comparisons with the dipole amplitude. }
\label{fig:DUST545MODEL} 
\end{figure}

Figure~\ref{fig:DUST545MODEL} shows the consistency of the two dust maps at 545\,GHz (top row).
The differences are shown in the bottom row for the full resolution, and at $5 \deg$ smoothing. Even if there are some differences near the galactic ridge, both maps are very consistent at high latitudes, demonstrating that the dust model stands at large scale and high latitude up to 857\,GHz.

\end{document}